\documentclass[%
print,
superscriptaddress,
 amsmath,amssymb,
 aps,twocolumn
]{revtex4}

\usepackage{times}
\usepackage{graphicx}
\usepackage{dcolumn}
\usepackage{bm}
\usepackage{dsfont}
\usepackage{mathrsfs}
\usepackage[landscape,papersize={297.1mm,210mm},left=1.9cm,right=1.6cm,top=2.7cm,bottom=2.8cm]{geometry}
\usepackage[                   
            pdfstartview=FitH,
            colorlinks, 
            pdfborder=001,   
            linkcolor=blue,
            anchorcolor=blue,
            citecolor=blue,
            urlcolor=blue
            ]{hyperref}
\usepackage{amsthm}
\makeatletter

\newcommand{\Rmnum}[1]{\expandafter\@slowromancap\romannumeral #1@}

\makeatother

\begin{document}
\title{Strong quantum nonlocality without entanglement in every $(n-1)$-partition}

\author{Huaqi Zhou}
\affiliation{School of Mathematics and Science, Hebei GEO University, Shijiazhuang 050031, China}
\author{Ting Gao}
\email{gaoting@hebtu.edu.cn}
\affiliation{School of Mathematical Sciences, Hebei Normal University, Shijiazhuang 050024, China}
\affiliation{Hebei Mathematics Research Center,  Hebei Normal University, Shijiazhuang 050024, China}
\affiliation{Hebei International Joint Research Center for Mathematics and Interdisciplinary Science, Hebei Normal University, Shijiazhuang 050024, China}
\author{Fengli Yan}
\email{flyan@hebtu.edu.cn}
\affiliation{College of Physics, Hebei Key Laboratory of Photophysics Research and Application, Hebei Normal University, Shijiazhuang 050024, China}

\begin{abstract}
Orthogonal product sets that are locally irreducible in every bipartition have the strongest nonlocality while also need a large number of quantum states. In this paper, we construct the orthogonal product sets with strong quantum nonlocality in any possible $n$-partite systems, where $n$ is greater than three. Rigorous proofs show that these sets are locally irreducible in every $(n-1)$-partition. They not only possess stronger properties than nonlocality and fewer quantum states than the strongest nonlocal sets, but also are positive answers to the open question `` how to construct different strength nonlocality of orthogonal product states for general multipartite and high-dimensional quantum systems" of Zhang et al. [\href{https://journals.aps.org/prl/abstract/10.1103/PhysRevA.99.062108} {Phys. Rev. A \textbf{99}, 062108 (2019)}]. Our results can also enhance one understanding for the nonlocality without entanglement.
~\\


\end{abstract}

\maketitle

\section{Introduction}
When a set of orthogonal quantum states is local indistinguishable, it reflects the fundamental feature of quantum mechanics, i.e. quantum nonlocality \cite{BennettDFMRSSW,Niset,HGY24,ZhangZGWO,Shi2022}. Here the local indistinguishability means that these orthogonal states in the set cannot be perfectly discriminated by local operations and classical communication (LOCC).
Different from Bell nonlocality which arises from entangled states \cite{Bell6466,Clauser69,Freedman72,Aspect82,Yan11,Gao14,Meng18,Chen04,DHYG15,DHYG16}, quantum nonlocality based on local indistinguishability is not limited by entangled states. There exists a special phenomenon of quantum nonlocality without entanglement.
An original result is the locally indistinguishable orthogonal product basis in $\mathcal{C}^{3}\otimes \mathcal{C}^{3}$ quantum system, which was presented by Bennett et al. in 1999 \cite{BennettDFMRSSW}. Later, in multipartite or high-dimensional quantum systems, many orthogonal product sets (OPSs) with quantum nonlocality were appeared successively \cite{Niset,ZhangGTCW,WangLZF,WLZF,Feng,Xu,ZhangZGWO,SHalder,Rout,Jiang,ZhenFZ,Rout21,Li21}.
This type of quantum nonlocality can also be used for quantum data hiding \cite{Terhal,DiVincenzo,Eggeling} and quantum secret sharing \cite{Hillery,Guo,Hsu,Markham,Rahaman,JWang} and show the meaningful research value.

Recently, the authors have increasingly utilized the local irreducibility to exhibit the quantum nonlocality. A set of orthogonal quantum states being locally irreducible refers to the inability to discriminate one or more quantum states from the set by orthogonality-preserving local measurements \cite{Halder}. Comparing with the local indistinguishability, local irreducibility can reveal stronger nonlocality. In 2019, Halder et al. \cite{Halder} proposed the concept of strong quantum nonlocality without entanglement. Specifically, in a composite quantum system $\otimes_{i=1}^{n}\mathcal{C}^{d_{i}}(n\geq 3,d_{i}\geq 3)$, an OPS is strongly nonlocal if it is locally irreducible in every bipartition \cite{Halder}. Here $\mathcal{C}^{d_{i}}$ stands for a $d_{i}$-dimensional Hilbert space. Actually, this is the strongest quantum nonlocality because an OPS is easier to distinguish locally as the more subsystems coming together. About this strongest nonlocality without entanglement, there are many achievements \cite{Halder,Yuan,Shi22,Che,He,ZGY,ZGY23}. Halder et al. \cite{Halder} and Yuan et al. \cite{Yuan} presented several OPSs with the strongest quantum nonlocal in three- and four-partite systems. He et al. \cite{He} gave the strongest quantum nonlocal OPSs in any possible $n$-partite systems for all odd $n$. We have also done some works in this area. In $\mathcal{C}^{d_{A}}\otimes \mathcal{C}^{d_{B}}\otimes \mathcal{C}^{d_{C}}$ $(d_{A},d_{B},d_{C}\geq 4)$ and $\mathcal{C}^{d_{A}}\otimes \mathcal{C}^{d_{B}}\otimes \mathcal{C}^{d_{C}}\otimes \mathcal{C}^{d_{D}}$ $(d_{A},d_{B},d_{C},d_{D}\geq 3)$ systems, we proposed the strongest nonlocal OPSs with fewer quantum states \cite{ZGY}. For any possible even party systems, we complemented the sets of orthogonal product states with the strongest quantum nonlocality \cite{ZGY23}. We and He et al. \cite{He} jointly constructed a general structure of incomplete orthogonal product bases with the strongest nonlocality in any possible multipartite and high-dimensional systems.

Of course, the OPSs with the strongest quantum nonlocality possess the best property. However, in $n$-qudit systems, these sets require $\Theta(d^{n-1})$ product states, which is a very vast quantity. So, it is necessary to consider the generally strong quantum nonlocality. Recently, a general definition of generally strong quantum nonlocality has been presented by Zhang et al. \cite{ZhangZ}, namely, in multipartite systems $\otimes_{i=1}^{n}\mathcal{C}^{d_{i}}(n\geq 3,d_{i}\geq 3)$, if an OPS is locally irreducible in every $(n-1)$-partition, then it is strongly nonlocal. Here $(n-1)$-partition represents the whole quantum system is divided into $n-1$ parts. In $\mathcal{C}^{3}\otimes \mathcal{C}^{3}\otimes \mathcal{C}^{3}\otimes \mathcal{C}^{3}$, they also provided an OPS which is locally irreducible in every tripartition but locally reducible in every bipartition. For the other possible systems, there are still unknowns regarding the structures of the OPS with generally strong rather than the strongest quantum nonlocality.

In this paper, we mainly study the generally strong quantum nonlocality without entanglement in every $(n-1)$-partition. We first introduce some necessary notations, definitions, and conclusions in Sec. \ref{Q2}. In Sec. \ref{Q3}, we present two structures of OPS with strong quantum nonlocality in four-partite systems, asymmetric and symmetric. Both of these new OPSs contain fewer quantum states than the OPS proposed by Zhang et al. \cite{ZhangZ}. In Sec. \ref{Q4}, we construct the strong nonlocal OPSs in odd party and even party systems, respectively. On the one hand, our OPSs require $\Theta(d^{\lceil\frac{n+1}{2}\rceil})$ product states which are less than the OPSs with the strongest nonlocality. On the other hand, the structures of these OPSs are concerned with arbitrary possible multipartite and high-dimensional quantum systems, it means that we answer the open question given by Zhang et al. \cite{ZhangZ} that how to construct different strength nonlocality of orthogonal product states for general multipartite and high-dimensional quantum systems. Finally, a brief summary as a end is presented in Sec. \ref{Q5}.

\section{Preliminaries}\label{Q2}
Given an $n$-partite quantum system $\mathcal{H}=\otimes_{k=1}^{n}\mathcal{C}^{d_{k}}$ $(n\geq3,d_{k}\geq3)$, let $\mathcal{B}=\{|i\rangle\}_{i=0}^{d_1d_2\cdots d_n -1}=\{\otimes_{k=1}^{n}|i_{k}\rangle ~ | ~ i_k=0, 1, \cdots, d_{k}-1 \}=\mathcal{B}^{\{1\}}\otimes \mathcal{B}^{\{2\}}\otimes \cdots\otimes \mathcal{B}^{\{n\}}$ denote the computational basis of the whole quantum system and $\mathcal{B}^{\{k\}}=\{|i_{k}\rangle\}_{{i_k}=0}^{d_{k}-1}$ is the computational basis of the $k$th subsystem.

\emph{Definition 1} \cite{ZhangZ}. In the $n$-partite quantum system $\mathcal{H}$, if an OPS is locally irreducible in every $(n-1)$-partition, then it is strongly nonlocal.

Unless otherwise specified, the strong nonlocality mentioned below is described by the Definition 1. It is well known that when all the POVM elements of a measurement are proportional to the identity operator, in other words, the measurement is trivial \cite{Walgate}, it produces no information from the resulting postmeasurement states. Therefore, the OPS in $\mathcal{H}$ must be locally irreducible if the orthogonality-preserving POVM performed on every subsystem can only be trivial \cite{Halder}. The converse is not sure in general.

Denote $A$ and $B$ as the subsets of $\{1,2,\ldots,n\}$, respectively. They are disjoint. $\overline{AB}$ expresses the complement of $A\cup B$. We have the following lemma.

\emph{Lemma~1}. For an OPS in the $n$-partite system $\mathcal{H}$, if any orthogonality-preserving POVM element performed on $AB$ party is only proportional to the identity operator, then any orthogonality-preserving POVM performed on the parties $A$ and $B$ can only be trivial, respectively.

$Proof.$ Let $\{|\psi_{i}\rangle=|\psi_{i}\rangle_{A}|\psi_{i}\rangle_{B}|\psi_{i}\rangle_{\overline{AB}}\}_{i=0}^{s}$ be an OPS in $n$-partite system $\mathcal{H}=\mathcal{H}_{A|B|\overline{AB}}$. The orthogonality-preserving POVM performed on the parties $AB$, $A$, and $B$ are denoted as $\{E_{AB}\}$, $\{E_{A}\}$, and $\{E_{B}\}$, respectively. Obviously, the POVM $\{E_{A}\otimes E_{B}\}$ is a special case of $\{E_{AB}\}$. If the conditions $\langle\psi_{i}|E_{AB}|\psi_{j}\rangle=0$ $(i\neq j)$ can deduce $E_{AB}\propto \mathbb{I}$, then $E_{A}$ and $E_{B}$ must be proportional to the identity operator. So, any orthogonality-preserving POVM performed on the parties $A$ and $B$ can only be trivial, respectively.
\hfill $\blacksquare$

Let $X_{kl}=\{k,l|k\neq l\}\subset \{1,2,\ldots,n\}$ be the union of any two subsystems of $\mathcal{H}$. According to the Lemma 1, we get a sufficient condition for the OPS being strong nonlocality.

\emph{Theorem 1}. The $n$-partite OPS is strongly nonlocal, if there always exists a subset in every bipartition $X_{kl}|\overline{X_{kl}}$ such that any orthogonality-preserving POVM performed on $X_{kl}$ party can only be trivial.

$Proof.$ Let $\{|\psi_{i}\rangle=|\psi_{i}\rangle_{X_{kl}}|\psi_{i}\rangle_{\overline{X_{kl}}}\}_{i=0}^{s}$ be an OPS in $n$-partite system $\mathcal{H}=\mathcal{H}_{X_{kl}|\overline{X_{kl}}}$. The orthogonality-preserving POVM performed on the party $X_{kl}$ is expressed as $\{E_{kl}\}$. If there exists a subset $\{|\psi_{i}\rangle=|\psi_{i}\rangle_{X_{kl}}|\psi_{i}\rangle_{\overline{X_{kl}}}\}_{i=i_{0}}^{i_{t}}$ $(\{i_{0},i_{1},\ldots,i_{t}\}\subset \{0,1,\ldots,s\})$ such that the conditions $\langle\psi_{i}|E_{kl}|\psi_{j}\rangle=0$ $(i,j\in \{i_{0},i_{1},\ldots,i_{t}\}$ and $i\neq j)$ can deduce $E_{kl}\propto \mathbb{I}$, then conditions $\langle\psi_{i}|E_{kl}|\psi_{j}\rangle=0$ $(i,j\in \{0,1,\ldots,s\}$ and $i\neq j)$ must deduce $E_{kl}\propto \mathbb{I}$. That is, for this OPS, any orthogonality-preserving POVM performed on the party $X_{kl}$ must be trivial. Based on the Lemma 1, we can know that the orthogonality-preserving POVM performed on the $k$th and $l$th subsystems can only be trivial, respectively. So, in every $(n-1)$-partition, the local orthogonality-preserving POVM performed on every subsystem can only be trivial. This means that the OPS is locally irreducible in every $(n-1)$-partition. Thus, the $n$-partite OPS is strongly nonlocal.
\hfill $\blacksquare$

It is worth noting that the subsets selected in diverse bipartitions can be different. Below, we will provide specific OPSs with strong nonlocality in multipartite systems. For simplicity, all product states are not normalized.

\section{The OPS with strong nonlocality in $4$-qudit system}\label{Q3}
\subsection{asymmetric}
According to the definitions of Refs. \cite{Halder,ZhangZ}, in three-partite system, if an OPS is strongly nonlocal, then it is also the strongest nonlocal. Our work starts from four-partite system. In $\mathcal{C}^{3}\otimes\mathcal{C}^{3}\otimes\mathcal{C}^{3}\otimes\mathcal{C}^{3}$, let
\begin{equation}\label{21}
\begin{aligned}
&S_{1}=\{|0\rangle|0\rangle|0\pm 1\rangle|\alpha_{i}\rangle\}_{i},\\
&S_{2}=\{|0\rangle|0\rangle|2\rangle|0\pm 1\rangle\},\\
&S_{3}=\{|0\pm 1\rangle|\alpha_{i}\rangle|2\rangle|2\rangle\}_{i},\\
&S_{4}=\{|2\rangle|0\pm 1\rangle|2\rangle|2\rangle\},\\
&S_{5}=\{|2\rangle|2\rangle|1\pm 2\rangle|\alpha_{i}\rangle\}_{i},\\
&S_{6}=\{|2\rangle|2\rangle|0\rangle|1\pm 2\rangle\},\\
&S_{7}=\{|1\pm 2\rangle|\alpha_{i}\rangle|0\rangle|0\rangle\}_{i},\\
&S_{8}=\{|0\rangle|1\pm 2\rangle|0\rangle|0\rangle\},\\
&S_{9}=\{|0\rangle|1\rangle|0\rangle|1\pm 2\rangle\},\\
&S_{10}=\{|2\rangle|1\rangle|2\rangle|0\pm 1\rangle\},\\
&S_{11}=\{|1\rangle|2\rangle|0\pm 1\rangle|2\rangle\},\\
&S_{12}=\{|1\rangle|0\rangle|1\pm 2\rangle|0\rangle\},\\
&S_{13}=\{|0\rangle|1\pm 2\rangle|1\rangle|2\rangle\},\\
&S_{14}=\{|2\rangle|0\pm 1\rangle|1\rangle|0\rangle\},\\
&S_{15}=\{|1\rangle|2\rangle|2\rangle|0\pm 1\rangle\},\\
&S_{16}=\{|1\rangle|0\rangle|0\rangle|1\pm 2\rangle\},\\
&S_{17}=\{|0\rangle|2\rangle|0\pm 2\rangle|1\rangle\},\\
&S_{18}=\{|2\rangle|0\rangle|0\pm 2\rangle|1\rangle\},\\
\end{aligned}
\end{equation}
where $|\alpha_{i}\rangle=\sum_{u=0}^{2}\omega_{3}^{iu}|u\rangle$ with $i=0,1,2$. For any positive integer $n$, here and below we use the notation $\omega_{n}:=\textup{e}^{\frac{2\pi\textup{i}}{n}}$. Note that, this OPS $\cup_{r=1}^{18}S_{r}$ (\ref{21}) is not invariant under cyclic permutation of the parties $\{1\}$, $\{2\}$, $\{3\}$, and $\{4\}$. It is asymmetric. The following theorem demonstrates that this OPS is strongly nonlocal.

\emph{Theorem 2}. In $\mathcal{C}^{3}\otimes\mathcal{C}^{3}\otimes\mathcal{C}^{3}\otimes\mathcal{C}^{3}$, the set $\cup_{r=1}^{18}S_{r}$ given by Eq. (\ref{21}) is strongly nonlocal in every three-partition. The size of this set is 52.

According to the Theorem 1, the issue of considering the strong quantum nonlocality of set $\cup_{r=1}^{18}S_{r}$ (\ref{21}) can be converted to consider the union orthogonality-preserving POVM performed on any two parties. Thus, we only need to discuss the orthogonality-preserving POVM performed on parties $X_{12}$, $X_{13}$, $X_{14}$, $X_{23}$, $X_{24}$, and $X_{34}$, respectively. Here, it is highly efficient by using the \cite[Theorem 1]{ZGY} to illustrate the triviality of orthogonality-preserving POVM. Appendixes \ref{A} and \ref{B} provide the specific explanation of \cite[Theorem 1]{ZGY} and the proof process of Theorem 2, respectively.

A strongly nonlocal OPS is more optimal when it contains fewer quantum states. On the one hand, we can achieve the same goal with fewer quantum resources. On the other hand, when we add some product states to a strongly nonlocal OPS, the new OPS remains the strong nonlocality. The reverse is generally not true. In $\mathcal{C}^{3}\otimes\mathcal{C}^{3}\otimes\mathcal{C}^{3}\otimes\mathcal{C}^{3}$, Zhang et al. \cite{ZhangZ} constructed the strongly nonlocal OPS containing 56 states. Our OPS (\ref{21}) has 52 quantum states, which is 4 elements fewer than OPS presented by Zhang et al.

Let $S_{r}^{[k]}=\{\textup{Tr}_{\bar{k}}(|\phi^{r}\rangle\langle\phi^{r} |)~|~|\phi^{r}\rangle\in S_{r}\}$ be the set of reduced density matrices and $S_{r}^{(k)}=\{|i\rangle_{k}\in\mathcal{B}^{\{k\}}|~\langle i|\phi^r\rangle_{k}\neq 0~\textup{for~any}~|\phi^r\rangle_{k}\in S_{r}^{[k]}\}$ represent the projection set of $S_{r}$ on the $k$th subsystem, where $k\in \{1,2,3,4\}$ and $r\in\{1,2,\ldots,18\}$. Example, $\{|0\rangle_{1}\}$, $\{|0\rangle_{2}\}$, $\{|0\rangle_{3},|1\rangle_{3}\}$, and $\{|0\rangle_{4},|1\rangle_{4},|2\rangle_{4}\}$ are the projection sets of $S_{1}$ on the 1th, 2th, 3th, and 4th subsystems, respectively. It is obvious that $S_{r}^{(k)}\subset\mathcal{B}^{\{k\}}$ and $S_{r}^{[k]}$ is a basis spanned by the corresponding computational basis $S_{r}^{(k)}$. Based on these characteristics, we extend the above structure (\ref{21}) to any finite dimensional system. For the projection set of every subset $S_{r}$ on the $k$th subsystem, the state $|0\rangle_{k}$ is invariant and $|1\rangle_{k}$ and $|2\rangle_{k}$ are replaced by $\{|1\rangle_{k},|2\rangle_{k},\ldots,|d_{k}-2\rangle_{k}\}$ and $\{|d_{k}-1\rangle_{k}\}$, respectively. Then, in system $\mathcal{C}^{d_{1}}\otimes\mathcal{C}^{d_{2}}\otimes\mathcal{C}^{d_{3}}\otimes\mathcal{C}^{d_{4}}$, we get
\begin{equation}\label{22}
\begin{aligned}
&H_{1}=\{|0\rangle_{1}|0\rangle_{2}|\beta_{j}\rangle_{3}|\alpha_{i}\rangle_{4}\}_{i,j},\\
&H_{2}=\{|0\rangle_{1}|0\rangle_{2}|d_{3}'\rangle_{3}|\beta_{j}\rangle_{4}\}_{j},\\
&H_{3}=\{|\beta_{j}\rangle_{1}|\alpha_{i}\rangle_{2}|d_{3}'\rangle_{3}|d_{4}'\rangle_{4}\}_{i,j},\\
&H_{4}=\{|d_{1}'\rangle_{1}|\beta_{j}\rangle_{2}|d_{3}'\rangle_{3}|d_{4}'\rangle_{4}\}_{j},\\
&H_{5}=\{|d_{1}'\rangle_{1}|d_{2}'\rangle_{2}|\gamma_{m}\rangle_{3}|\alpha_{i}\rangle_{4}\}_{i,m},\\
&H_{6}=\{|d_{1}'\rangle_{1}|d_{2}'\rangle_{2}|0\rangle_{3}|\gamma_{m}\rangle_{4}\}_{m},\\
&H_{7}=\{|\gamma_{m}\rangle_{1}|\alpha_{i}\rangle_{2}|0\rangle_{3}|0\rangle_{4}\}_{i,m},\\
&H_{8}=\{|0\rangle_{1}|\gamma_{m}\rangle_{2}|0\rangle_{3}|0\rangle_{4}\}_{m},\\
&H_{9}=\{|0\rangle_{1}|\kappa_{l}\rangle_{2}|0\rangle_{3}|\gamma_{m}\rangle_{4}\}_{l,m},\\
&H_{10}=\{|d_{1}'\rangle_{1}|\kappa_{l}\rangle_{2}|d_{3}'\rangle_{3}|\beta_{j}\rangle_{4}\}_{j,l},\\
&H_{11}=\{|\kappa_{l}\rangle_{1}|d_{2}'\rangle_{2}|\beta_{j}\rangle_{3}|d_{4}'\rangle_{4}\}_{j,l},\\
&H_{12}=\{|\kappa_{l}\rangle_{1}|0\rangle_{2}|\gamma_{m}\rangle_{3}|0\rangle_{4}\}_{l,m},\\
&H_{13}=\{|0\rangle_{1}|\gamma_{m}\rangle_{2}|\kappa_{l}\rangle_{3}|d_{4}'\rangle_{4}\}_{l,m},\\
&H_{14}=\{|d_{1}'\rangle_{1}|\beta_{j}\rangle_{2}|\kappa_{l}\rangle_{3}|0\rangle_{4}\}_{j,l},\\
\end{aligned}
\end{equation}
\begin{equation*}
\begin{aligned}
&H_{15}=\{|\kappa_{l}\rangle_{1}|d_{2}'\rangle_{2}|d_{3}'\rangle_{3}|\beta_{j}\rangle_{4}\}_{j,l},\\
&H_{16}=\{|\kappa_{l}\rangle_{1}|0\rangle_{2}|0\rangle_{3}|\gamma_{m}\rangle_{4}\}_{l,m},\\
&H_{17}=\{|0\rangle_{1}|d_{2}'\rangle_{2}|0\pm d_{3}'\rangle_{3}|\kappa_{l}\rangle_{4}\}_{l},\\
&H_{18}=\{|d_{1}'\rangle_{1}|0\rangle_{2}|0\pm d_{3}'\rangle_{3}|\kappa_{l}\rangle_{4}\}_{l},\\
\end{aligned}
\end{equation*}
where $|\alpha_{i}\rangle_{k}=\sum_{u=0}^{d_{k}-1}\omega_{d_{k}}^{iu}|u\rangle$, $|\beta_{j}\rangle_{k}=\sum_{u=0}^{d_{k}-2}\omega_{d_{k}-1}^{ju}|u\rangle$, $|\gamma_{m}\rangle_{k}=\sum_{u=0}^{d_{k}-2}\omega_{d_{k}-1}^{mu}|u+1\rangle$, $|\kappa_{l}\rangle_{k}=\sum_{u=0}^{d_{k}-3}\omega_{d_{k}-2}^{lu}|u+1\rangle$, $d_{k}'=d_{k}-1$ for $i\in \mathcal{Z}_{d_{k}}$, $j,m\in \mathcal{Z}_{d_{k}-1}$, $l\in \mathcal{Z}_{d_{k}-2}$, and $k\in\{1,2,3,4\}$. We find that it is still strongly nonlocal.

\emph{Theorem 3}. In $\mathcal{C}^{d_{1}}\otimes\mathcal{C}^{d_{2}}\otimes\mathcal{C}^{d_{3}}\otimes\mathcal{C}^{d_{4}}$, the set $\cup_{r=1}^{18}H_{r}$ given by Eq. (\ref{22}) is strongly nonlocal in every three-partition.

The specific proof is given in Appendix \ref{C}. When $d_{1}=d_{2}=d_{3}=d_{4}=d$, the size of $\cup_{r=1}^{18}H_{r}$ (\ref{22}) is $12d^2-20d+4$. Compared with the strongest nonlocal OPS in Ref. \cite{ZGY}, which has $8d^3-24d^2+32d-18$, the OPS of Theorem 3 requires fewer quantum states, although the nonlocality of this OPS is weaker.

\subsection{symmetric}
In this subsection, we will give an OPS with symmetric structure, i.e., this OPS is invariant under cyclic permutation of all parties. In the low-dimensional system $\mathcal{C}^{3}\otimes\mathcal{C}^{3}\otimes\mathcal{C}^{3}\otimes\mathcal{C}^{3}$, let
\begin{equation}\label{23}
\begin{aligned}
&S_{1,1}=\{|1\rangle|1\rangle|0\rangle|1\pm 2\rangle\},\\
&S_{1,2}=\{|1\rangle|0\rangle|1\pm 2\rangle|1\rangle\},\\
&S_{1,3}=\{|0\rangle|1\pm 2\rangle|1\rangle|1\rangle\},\\
&S_{1,4}=\{|1\pm 2\rangle|1\rangle|1\rangle|0\rangle\},\\
&S_{2,1}=\{|2\rangle|0\rangle|0\rangle|0\pm 1\rangle\},\\
&S_{2,2}=\{|0\rangle|0\rangle|0\pm 1\rangle|2\rangle\},\\
&S_{2,3}=\{|0\rangle|0\pm 1\rangle|2\rangle|0\rangle\},\\
&S_{2,4}=\{|0\pm 1\rangle|2\rangle|0\rangle|0\rangle\},\\
&S_{3,1}=\{|0\rangle|0\rangle|2\rangle|1\pm 2\rangle\},\\
&S_{3,2}=\{|0\rangle|2\rangle|1\pm 2\rangle|0\rangle\},\\
&S_{3,3}=\{|2\rangle|1\pm 2\rangle|0\rangle|0\rangle\},\\
&S_{3,4}=\{|1\pm 2\rangle|0\rangle|0\rangle|2\rangle\},\\
&S_{4,1}=\{|1\rangle|2\rangle|2\rangle|0\pm 2\rangle\},\\
&S_{4,2}=\{|2\rangle|2\rangle|0\pm 2\rangle|1\rangle\},\\
&S_{4,3}=\{|2\rangle|0\pm 2\rangle|1\rangle|2\rangle\},\\
&S_{4,4}=\{|0\pm 2\rangle|1\rangle|2\rangle|2\rangle\},\\
&S_{5,1}=\{|2\rangle|2\rangle|1\rangle|0\pm 1\rangle\},\\
&S_{5,2}=\{|2\rangle|1\rangle|0\pm 1\rangle|2\rangle\},\\
&S_{5,3}=\{|1\rangle|0\pm 1\rangle|2\rangle|2\rangle\},\\
&S_{5,4}=\{|0\pm 1\rangle|2\rangle|2\rangle|1\rangle\},\\
&S_{6,1}=\{|0\rangle|2\rangle|2\rangle|2\rangle\},\\
&S_{6,2}=\{|2\rangle|2\rangle|2\rangle|0\rangle\},\\
&S_{6,3}=\{|2\rangle|2\rangle|0\rangle|2\rangle\},\\
&S_{6,4}=\{|2\rangle|0\rangle|2\rangle|2\rangle\},\\
&S_{7,1}=\{|0\rangle|1\rangle|2\rangle|1\rangle\},\\
&S_{7,2}=\{|1\rangle|2\rangle|1\rangle|0\rangle\},\\
\end{aligned}
\end{equation}
\begin{equation*}
\begin{aligned}
&S_{7,3}=\{|2\rangle|1\rangle|0\rangle|1\rangle\},\\
&S_{7,4}=\{|1\rangle|0\rangle|1\rangle|2\rangle\},\\
&S_{8,1}=\{|2\rangle|1\rangle|2\rangle|1\rangle\},\\
&S_{8,2}=\{|1\rangle|2\rangle|1\rangle|2\rangle\}.\\
\end{aligned}
\end{equation*}

\emph{Theorem 4}. In $\mathcal{C}^{3}\otimes\mathcal{C}^{3}\otimes\mathcal{C}^{3}\otimes\mathcal{C}^{3}$, the set $\cup_{r,t}S_{r,t}$ given by Eq. (\ref{23}) is strongly nonlocal in every three-partition. The size of this OPS is 50.

Due to the symmetry, parties $X_{12}$, $X_{23}$, $X_{34}$, and $X_{14}$ have the same situation, and the situation of the $X_{13}$ party is same as the $X_{24}$ party. So, we only need to consider the orthogonality-preserving POVM performed on parties $X_{34}$ and $X_{24}$, respectively. The detailed proof is shown in Appendix \ref{D}.

By using the same method, we convert $|1\rangle_{k}$ and $|2\rangle_{k}$ in the projection set $S_{r,t}^{(k)}$ to $\{|1\rangle_{k},|2\rangle_{k},\ldots,|d_{k}-2\rangle_{k}\}$ and $\{|d_{k}-1\rangle_{k}\}$, respectively. Then, the structure of OPS in Theorem 4 can be generalized to high-dimensional system. In $\mathcal{C}^{d_{1}}\otimes\mathcal{C}^{d_{2}}\otimes\mathcal{C}^{d_{3}}\otimes\mathcal{C}^{d_{4}}$, we have
\begin{equation}\label{24}
\begin{aligned}
&H_{1,1}=\{|\kappa_{l}\rangle_{1}|\kappa_{l}\rangle_{2}|0\rangle_{3}|\gamma_{m}\rangle_{4}\}_{m,l|_{1},l|_{2}},\\
&H_{1,2}=\{|\kappa_{l}\rangle_{1}|0\rangle_{2}|\gamma_{m}\rangle_{3}|\kappa_{l}\rangle_{4}\}_{m,l|_{1},l|_{4}},\\
&H_{1,3}=\{|0\rangle_{1}|\gamma_{m}\rangle_{2}|\kappa_{l}\rangle_{3}|\kappa_{l}\rangle_{4}\}_{m,l|_{3},l|_{4}},\\
&H_{1,4}=\{|\gamma_{m}\rangle_{1}|\kappa_{l}\rangle_{2}|\kappa_{l}\rangle_{3}|0\rangle_{4}\}_{m,l|_{2},l|_{3}},\\
&H_{2,1}=\{|d_{1}'\rangle_{1}|0\rangle_{2}|0\rangle_{3}|\beta_{j}\rangle_{4}\}_{j},\\
&H_{2,2}=\{|0\rangle_{1}|0\rangle_{2}|\beta_{j}\rangle_{3}|d_{4}'\rangle_{4}\}_{j},\\
&H_{2,3}=\{|0\rangle_{1}|\beta_{j}\rangle_{2}|d_{3}'\rangle_{3}|0\rangle_{4}\}_{j},\\
&H_{2,4}=\{|\beta_{j}\rangle_{1}|d_{2}'\rangle_{2}|0\rangle_{3}|0\rangle_{4}\}_{j},\\
&H_{3,1}=\{|0\rangle_{1}|0\rangle_{2}|d_{3}'\rangle_{3}|\gamma_{m}\rangle_{4}\}_{m},\\
&H_{3,2}=\{|0\rangle_{1}|d_{2}'\rangle_{2}|\gamma_{m}\rangle_{3}|0\rangle_{4}\}_{m},\\
&H_{3,3}=\{|d_{1}'\rangle_{1}|\gamma_{m}\rangle_{2}|0\rangle_{3}|0\rangle_{4}\}_{m},\\
&H_{3,4}=\{|\gamma_{m}\rangle_{1}|0\rangle_{2}|0\rangle_{3}|d_{4}'\rangle_{4}\}_{m},\\
&H_{4,1}=\{|\kappa_{l}\rangle_{1}|d_{2}'\rangle_{2}|d_{3}'\rangle_{3}|0\pm d_{4}'\rangle_{4}\}_{l},\\
&H_{4,2}=\{|d_{1}'\rangle_{1}|d_{2}'\rangle_{2}|0\pm d_{3}'\rangle_{3}|\kappa_{l}\rangle_{4}\}_{l},\\
&H_{4,3}=\{|d_{1}'\rangle_{1}|0\pm d_{2}'\rangle_{2}|\kappa_{l}\rangle_{3}|d_{4}'\rangle_{4}\}_{l},\\
&H_{4,4}=\{|0\pm d_{1}'\rangle_{1}|\kappa_{l}\rangle_{2}|d_{3}'\rangle_{3}|d_{4}'\rangle_{4}\}_{l},\\
&H_{5,1}=\{|d_{1}'\rangle_{1}|d_{2}'\rangle_{2}|\kappa_{l}\rangle_{3}|\beta_{j}\rangle_{4}\}_{j,l},\\
&H_{5,2}=\{|d_{1}'\rangle_{1}|\kappa_{l}\rangle_{2}|\beta_{j}\rangle_{3}|d_{4}'\rangle_{4}\}_{j,l},\\
&H_{5,3}=\{|\kappa_{l}\rangle_{1}|\beta_{j}\rangle_{2}|d_{3}'\rangle_{3}|d_{4}'\rangle_{4}\}_{j,l},\\
&H_{5,4}=\{|\beta_{j}\rangle_{1}|d_{2}'\rangle_{2}|d_{3}'\rangle_{3}|\kappa_{l}\rangle_{4}\}_{j,l},\\
&H_{6,1}=\{|0\rangle_{1}|d_{2}'\rangle_{2}|d_{3}'\rangle_{3}|d_{4}'\rangle_{4}\},\\
&H_{6,2}=\{|d_{1}'\rangle_{1}|d_{2}'\rangle_{2}|d_{3}'\rangle_{3}|0\rangle_{4}\},\\
&H_{6,3}=\{|d_{1}'\rangle_{1}|d_{2}'\rangle_{2}|0\rangle_{3}|d_{4}'\rangle_{4}\},\\
&H_{6,4}=\{|d_{1}'\rangle_{1}|0\rangle_{2}|d_{3}'\rangle_{3}|d_{4}'\rangle_{4}\},\\
&H_{7,1}=\{|0\rangle_{1}|\kappa_{l}\rangle_{2}|d_{3}'\rangle_{3}|\kappa_{l}\rangle_{4}\}_{l|_{2},l|_{4}},\\
&H_{7,2}=\{|\kappa_{l}\rangle_{1}|d_{2}'\rangle_{2}|\kappa_{l}\rangle_{3}|0\rangle_{4}\}_{l|_{1},l|_{3}},\\
&H_{7,3}=\{|d_{1}'\rangle_{1}|\kappa_{l}\rangle_{2}|0\rangle_{3}|\kappa_{l}\rangle_{4}\}_{l|_{2},l|_{4}},\\
&H_{7,4}=\{|\kappa_{l}\rangle_{1}|0\rangle_{2}|\kappa_{l}\rangle_{3}|d_{4}'\rangle_{4}\}_{l|_{1},l|_{3}},\\
&H_{8,1}=\{|d_{1}'\rangle_{1}|\kappa_{l}\rangle_{2}|d_{3}'\rangle_{3}|\kappa_{l}\rangle_{4}\}_{l|_{2},l|_{4}},\\
&H_{8,2}=\{|\kappa_{l}\rangle_{1}|d_{2}'\rangle_{2}|\kappa_{l}\rangle_{3}|d_{4}'\rangle_{4}\}_{l|_{1},l|_{3}}.\\
\end{aligned}
\end{equation}
Here $|\beta_{j}\rangle_{k}=\sum_{u=0}^{d_{k}-2}\omega_{d_{k}-1}^{ju}|u\rangle$, $|\gamma_{m}\rangle_{k}=\sum_{u=0}^{d_{k}-2}\omega_{d_{k}-1}^{mu}|u+1\rangle$, $|\kappa_{l}\rangle_{k}=\sum_{u=0}^{d_{k}-3}\omega_{d_{k}-2}^{lu}|u+1\rangle$, $d_{k}'=d_{k}-1$ for $i\in \mathcal{Z}_{d_{k}}$, $j,m\in \mathcal{Z}_{d_{k}-1}$, $l\in \mathcal{Z}_{d_{k}-2}$, and $k\in\{1,2,3,4\}$.

\emph{Theorem 5}. In $\mathcal{C}^{d_{1}}\otimes\mathcal{C}^{d_{2}}\otimes\mathcal{C}^{d_{3}}\otimes\mathcal{C}^{d_{4}}$, the set $\cup_{r,t}H_{r,t}$ given by Eq. (\ref{24}) has strong quantum nonlocality in every three-partition.

The specific proof is shown in Appendix \ref{E}. When $d_{1}=d_{2}=d_{3}=d_{4}=d$, this OPS (\ref{24}) contains $4d^3-10d^2+12d-4$ quantum states. When $d=3$, the set $\cup_{r,t}H_{r,t}$ (\ref{24}) is the OPS in Theorem 4, which has 50 quantum states. Meanwhile, it has 2 states fewer than the set $\cup_{r=1}^{18}S_{r}$ (\ref{21}). When $d>3$, it is obvious that the set $\cup_{r,t}H_{r,t}$ (\ref{24}) contains more quantum states than the set $\cup_{r=1}^{18}H_{r}$ in Theorem 3. Nevertheless, compared with $8d^3-24d^2+32d-18$ of the strongest nonlocal OPS in Ref. \cite{ZGY}, the OPS (\ref{24}) still requires fewer quantum states, specifically $4d^3-14d^2+20d-14$ fewer.

\section{Strong quantum nonlocality in every $(n-1)$-partition}\label{Q4}
Although symmetric structures may consume more quantum resources, their structures are simpler and the number of situations that need to be considered in the proof process is significantly reduced. Next, we will discuss the strongly nonlocal OPSs with symmetric structure in $n$-partite systems. Here $n$ is divided into two situations about whether $n$ is odd or even.
\subsection{$n$ is odd}
In system $\otimes_{k=1}^{n}\mathcal{C}^{d_{k}}$ ($n$ is odd, $n\geq 5$, and $d_{k}\geq 3$), we propose an OPS as shown below.

$H_{p1,1}^{O},H_{p1,2}^{O},\ldots,H_{p1,\frac{n-1}{2}}^{O},H_{p1,\frac{n+1}{2}}^{O},H_{p1,\frac{n+3}{2}}^{O},\ldots,H_{p1,n}^{O}$ are defined as
\begin{equation}\label{25}
\begin{aligned}
&\{|\alpha_{i}\rangle_{1}\cdots|\alpha_{i}\rangle_{\frac{n-1}{2}}|p\rangle_{\frac{n+1}{2}}|0\rangle_{\frac{n+3}{2}}\cdots|0\rangle_{n}\}_{i|_{1},\ldots,i|_{\frac{n-1}{2}}},\\
&\{|\alpha_{i}\rangle_{1}\cdots|p\rangle_{\frac{n-1}{2}}|0\rangle_{\frac{n+1}{2}}|0\rangle_{\frac{n+3}{2}}\cdots|\alpha_{i}\rangle_{n}\}_{i|_{n},\ldots,i|_{\frac{n-3}{2}}},\\
&\quad\vdots\\
&\{|\alpha_{i}\rangle_{1}|p\rangle_{2}\cdots|0\rangle_{\frac{n+3}{2}}|\alpha_{i}\rangle_{\frac{n+5}{2}}\cdots|\alpha_{i}\rangle_{n}\}_{i|_{\frac{n+5}{2}},\ldots,i|_{1}},\\
&\{|p\rangle_{1}|0\rangle_{2}\cdots|0\rangle_{\frac{n+1}{2}}|\alpha_{i}\rangle_{\frac{n+3}{2}}\cdots|\alpha_{i}\rangle_{n}\}_{i|_{\frac{n+3}{2}},\ldots,i|_{n}},\\
&\{|0\rangle_{1}\cdots|0\rangle_{\frac{n-1}{2}}|\alpha_{i}\rangle_{\frac{n+1}{2}}\cdots|\alpha_{i}\rangle_{n-1}|p\rangle_{n}\}_{i|_{\frac{n+1}{2}},\ldots,i|_{n-1}},\\
&\quad\vdots\\
&\{|0\rangle_{1}|\alpha_{i}\rangle_{2}\cdots|\alpha_{i}\rangle_{\frac{n+1}{2}}|p\rangle_{\frac{n+3}{2}}\cdots|0\rangle_{n}\}_{i|_{2},\ldots,i|_{\frac{n+1}{2}}},\\
\end{aligned}
\end{equation}
respectively. Here $|\alpha_{i}\rangle_{k}=\sum_{u=0}^{d_{k}-1}\omega_{d_{k}}^{iu}|u\rangle$ and $|p\rangle_{k}=|1\rangle_{k},|2\rangle_{k},\ldots,|d_{k}-1\rangle_{k}$ for $i\in \mathcal{Z}_{d_{k}}$ and $k\in\{1,2,\ldots,n\}$.

$H_{p2,1}^{O},H_{p2,2}^{O},\ldots,H_{p2,\frac{n-1}{2}}^{O},H_{p2,\frac{n+1}{2}}^{O},H_{p2,\frac{n+3}{2}}^{O},\ldots,H_{p2,n}^{O}$ are described as
\begin{equation}\label{26}
\begin{aligned}
&\{|\gamma_{m}\rangle_{1}|0\rangle_{2}\cdots|0\rangle_{\frac{n-1}{2}}|p\rangle_{\frac{n+1}{2}}|0\rangle_{\frac{n+3}{2}}\cdots|0\rangle_{n-1}|1\rangle_{n}\}_{m|_{1}},\\
&\{|0\rangle_{1}\cdots|0\rangle_{\frac{n-3}{2}}|p\rangle_{\frac{n-1}{2}}|0\rangle_{\frac{n+1}{2}}\cdots|0\rangle_{n-2}|1\rangle_{n-1}|\gamma_{m}\rangle_{n}\}_{m|_{n}},\\
&\quad\vdots\\
&\{|0\rangle_{1}|p\rangle_{2}|0\rangle_{3}\cdots|0\rangle_{\frac{n+1}{2}}|1\rangle_{\frac{n+3}{2}}|\gamma_{m}\rangle_{\frac{n+5}{2}}\cdots|0\rangle_{n}\}_{m|_{\frac{n+5}{2}}},\\
&\{|p\rangle_{1}|0\rangle_{2}\cdots|0\rangle_{\frac{n-1}{2}}|1\rangle_{\frac{n+1}{2}}|\gamma_{m}\rangle_{\frac{n+3}{2}}\cdots|0\rangle_{n}\}_{m|_{\frac{n+3}{2}}},\\
&\{|0\rangle_{1}\cdots|0\rangle_{\frac{n-3}{2}}|1\rangle_{\frac{n-1}{2}}|\gamma_{m}\rangle_{\frac{n+1}{2}}\cdots|0\rangle_{n-1}|p\rangle_{n}\}_{m|_{\frac{n+1}{2}}},\\
&\quad\vdots\\
&\{|1\rangle_{1}|\gamma_{m}\rangle_{2}\cdots|0\rangle_{\frac{n+1}{2}}|p\rangle_{\frac{n+3}{2}}|0\rangle_{\frac{n+5}{2}}\cdots|0\rangle_{n}\}_{m|_{2}},\\
\end{aligned}
\end{equation}
respectively. Here $|\gamma_{m}\rangle_{k}=\sum_{u=0}^{d_{k}-2}\omega_{d_{k}-1}^{mu}|u+1\rangle$ and $|p\rangle_{k}=|1\rangle_{k},|2\rangle_{k},\ldots,|d_{k}-1\rangle_{k}$ for $m\in \mathcal{Z}_{d_{k}-1}$ and $k\in\{1,2,\ldots,n\}$.

For the union of Eqs. (\ref{25}) and (\ref{26}), we have the following theorem.

\emph{Theorem 6}. In $\otimes_{k=1}^{n}\mathcal{C}^{d_{k}}$ ($n$ is odd, $n\geq 5$, and $d_{k}\geq 3$), the set $\cup_{p,r,t}H_{pr,t}^{O}$ given by Eqs. (\ref{25}) and (\ref{26}) has strong quantum nonlocality in every ($n$-1)-partition.

Combining Theorem 1 and Lemma 1, we only need to demonstrate the triviality of the orthogonality-preserving POVM performed on party $X_{1l}$ with $l=2,\ldots,\frac{n+1}{2}$. The specific process is shown in Appendix \ref{F}.

\subsection{$n$ is even}
In system $\otimes_{k=1}^{n}\mathcal{C}^{d_{k}}$ ($n$ is even, $n\geq 6$, and $d_{k}\geq 3$), let $H_{p1,1}^{E},H_{p1,2}^{E},\ldots,H_{p1,\frac{n}{2}-1}^{E},H_{p1,\frac{n}{2}}^{E},H_{p1,\frac{n}{2}+1}^{E},$ $H_{p1,\frac{n}{2}+2}^{E},\ldots,H_{p1,n}^{E}$ be equal to

\begin{widetext}
\begin{equation}\label{27}
\begin{aligned}
&\{|\alpha_{i}\rangle_{1}\cdots|\alpha_{i}\rangle_{\frac{n}{2}-1}|\gamma_{m}\rangle_{\frac{n}{2}}|p\rangle_{\frac{n}{2}+1}|0\rangle_{\frac{n}{2}+2}\cdots|0\rangle_{n}\}_{i|_{1},\ldots,i|_{\frac{n}{2}-1},m},\\
&\{|\alpha_{i}\rangle_{1}\cdots|\gamma_{m}\rangle_{\frac{n}{2}-1}|p\rangle_{\frac{n}{2}}|0\rangle_{\frac{n}{2}+1}\cdots|0\rangle_{n-1}|\alpha_{i}\rangle_{n}\}_{i|_{n},\ldots,i|_{\frac{n}{2}-2},m},\\
&\quad\vdots\\
&\{|\alpha_{i}\rangle_{1}|\gamma_{m}\rangle_{2}|p\rangle_{3}\cdots|0\rangle_{\frac{n}{2}+2}|\alpha_{i}\rangle_{\frac{n}{2}+3}\cdots|\alpha_{i}\rangle_{n}\}_{i|_{\frac{n}{2}+3},\ldots,i|_{1},m},\\
&\{|\gamma_{m}\rangle_{1}|p\rangle_{2}|0\rangle_{3}\cdots|0\rangle_{\frac{n}{2}+1}|\alpha_{i}\rangle_{\frac{n}{2}+2}\cdots|\alpha_{i}\rangle_{n}\}_{i|_{\frac{n}{2}+2},\ldots,i|_{n},m},\\
&\{|p\rangle_{1}|0\rangle_{2}\cdots|0\rangle_{\frac{n}{2}}|\alpha_{i}\rangle_{\frac{n}{2}+1}\cdots|\alpha_{i}\rangle_{n-1}|\gamma_{m}\rangle_{n}\}_{i|_{\frac{n}{2}+1},\ldots,i|_{n-1},m},\\
&\{|0\rangle_{1}\cdots|0\rangle_{\frac{n}{2}-1}|\alpha_{i}\rangle_{\frac{n}{2}}\cdots|\alpha_{i}\rangle_{n-2}|\gamma_{m}\rangle_{n-1}|p\rangle_{n}\}_{i|_{\frac{n}{2}},\ldots,i|_{n-2},m},\\
&\quad\vdots\\
&\{|0\rangle_{1}|\alpha_{i}\rangle_{2}\cdots|\alpha_{i}\rangle_{\frac{n}{2}}|\gamma_{m}\rangle_{\frac{n}{2}+1}|p\rangle_{\frac{n}{2}+2}\cdots|0\rangle_{n}\}_{i|_{2},\ldots,i|_{\frac{n}{2}},m},\\
\end{aligned}
\end{equation}
\end{widetext}
respectively. Here $|\alpha_{i}\rangle_{k}=\sum_{u=0}^{d_{k}-1}\omega_{d_{k}}^{iu}|u\rangle$, $|\gamma_{m}\rangle_{k}=\sum_{u=0}^{d_{k}-2}\omega_{d_{k}-1}^{mu}|u+1\rangle$, and $|p\rangle_{k}=|1\rangle_{k},|2\rangle_{k},\ldots,|d_{k}-1\rangle_{k}$ for $i\in \mathcal{Z}_{d_{k}}$, $m\in \mathcal{Z}_{d_{k}-1}$, and $k\in\{1,2,\ldots,n\}$.

$H_{p2,1}^{E},H_{p2,2}^{E},\ldots,H_{p2,n-1}^{E},H_{p2,n}^{E}$ are defined as
\begin{equation}\label{28}
\begin{aligned}
&\{|0\rangle_{1}|p\rangle_{2}|0\rangle_{3}\cdots|0\rangle_{n-2}|2\rangle_{n-1}|2\rangle_{n}\},\\
&\{|p\rangle_{1}|0\rangle_{2}\cdots|0\rangle_{n-3}|2\rangle_{n-2}|2\rangle_{n-1}|0\rangle_{n}\},\\
&\quad\vdots\\
&\{|2\rangle_{1}|2\rangle_{2}|0\rangle_{3}|p\rangle_{4}|0\rangle_{5}\cdots|0\rangle_{n}\},\\
&\{|2\rangle_{1}|0\rangle_{2}|p\rangle_{3}|0\rangle_{4}\cdots|0\rangle_{n-1}|2\rangle_{n}\},\\
\end{aligned}
\end{equation}
respectively. Here $|p\rangle_{k}=|1\rangle_{k},|2\rangle_{k},\ldots,|d_{k}-1\rangle_{k}$ for $k\in\{1,2,\ldots,n\}$.

$H_{p3,1}^{E},H_{p3,2}^{E},\ldots,H_{p3,n}^{E}$ are denoted as
\begin{equation}\label{29}
\begin{aligned}
&\{|\gamma_{m}\rangle_{1}|0\rangle_{2}\cdots|0\rangle_{\frac{n}{2}-1}|p\rangle_{\frac{n}{2}}|1\rangle_{\frac{n}{2}+1}\cdots|1\rangle_{n}\}_{m},\\
&\{|0\rangle_{1}\cdots|0\rangle_{\frac{n}{2}-2}|p\rangle_{\frac{n}{2}-1}|1\rangle_{\frac{n}{2}}\cdots|1\rangle_{n-1}|\gamma_{m}\rangle_{n}\}_{m},\\
&\quad\vdots\\
&\{|1\rangle_{1}|\gamma_{m}\rangle_{2}|0\rangle_{3}\cdots|0\rangle_{\frac{n}{2}}|p\rangle_{\frac{n}{2}+1}|1\rangle_{\frac{n}{2}+2}\cdots|1\rangle_{n}\}_{m},\\
\end{aligned}
\end{equation}
respectively. Here $|\gamma_{m}\rangle_{k}=\sum_{u=0}^{d_{k}-2}\omega_{d_{k}-1}^{mu}|u+1\rangle$ and $|p\rangle_{k}=|1\rangle_{k},|2\rangle_{k},\ldots,|d_{k}-1\rangle_{k}$ for $m\in \mathcal{Z}_{d_{k}-1}$ and $k\in\{1,2,\ldots,n\}$.

$H_{4,1}^{E},H_{4,2}^{E},\ldots,H_{4,n}^{E}$ are described as
\begin{equation}\label{30}
\begin{aligned}
&\{|0\pm 1\rangle_{1}|1\rangle_{2}|0\rangle_{3}\cdots|0\rangle_{\frac{n}{2}}|1\rangle_{\frac{n}{2}+1}|0\rangle_{\frac{n}{2}+2}\cdots|0\rangle_{n}\},\\
&\{|1\rangle_{1}|0\rangle_{2}\cdots|0\rangle_{\frac{n}{2}-1}|1\rangle_{\frac{n}{2}}|0\rangle_{\frac{n}{2}+1}\cdots|0\rangle_{n-1}|0\pm 1\rangle_{n}\},\\
&\quad\vdots\\
&\{|0\rangle_{1}|0\pm 1\rangle_{2}|1\rangle_{3}|0\rangle_{4}\cdots|0\rangle_{\frac{n}{2}+1}|1\rangle_{\frac{n}{2}+2}|0\rangle_{\frac{n}{2}+3}\cdots|0\rangle_{n}\},\\
\end{aligned}
\end{equation}
respectively.

The union of Eqs. (\ref{27})-(\ref{30}) satisfies the following theorem.

\emph{Theorem 7}. In $\otimes_{k=1}^{n}\mathcal{C}^{d_{k}}$ ($n$ is even, $n\geq 6$, and $d_{k}\geq 3$), the set $\cup_{p,r,t}H_{pr,t}^{E}$ given by Eqs. (\ref{27})-(\ref{30}) has strong quantum nonlocality in every ($n$-1)-partition.

Similar to the proof of Theorem 6, we only need to consider the orthogonality-preserving POVM performed on party $X_{1l}$ with $l=2,\ldots,\frac{n}{2}+1$. The detailed proof is shown in Appendix \ref{G}.

In Ref. \cite{Jiang}, Jiang et al. provided some nonlocal OPSs which are locally irreducible in every $n$-partition, but locally reducible in arbitrary ($n$-1)-partition. Compare with these OPSs, our results in Theorems 6 and 7 have stronger quantum nonlocality. In addition, when $d_{1}=d_{2}=\cdots=d_{n}=d$, the OPSs $\cup_{p,r,t}H_{pr,t}^{O}$ (\ref{25})-(\ref{26}) and $\cup_{p,r,t}H_{pr,t}^{E}$ (\ref{27})-(\ref{30}) contain $\Theta(d^{\lceil\frac{n+1}{2}\rceil})$ quantum states. The strongest nonlocal OPSs in Refs. \cite{He,ZGY23} need $\Theta(d^{n-1})$ quantum states which far exceeds the quantity required by our OPSs.

\section{Conclusion}\label{Q5}
We show the direct methods to construct multipartite orthogonal product sets with the strong quantum nonlocality in every ($n$-1)-partition. For the open question `` how to construct different strength nonlocality of orthogonal product states for general multipartite and high-dimensional quantum systems" of Zhang et al. \cite{ZhangZ}, we provide a positive answer. The specific results are presented by the OPSs in Theorems 2-7. They have two characteristics, namely, possessing stronger property than nonlocal OPSs and containing fewer quantum states than the strongest nonlocal OPSs. Our results not only can lead to a better understanding of the quantum nonlocality without entanglement, but also provide theoretical basis for people to design quantum protocols.

\begin{acknowledgments}
This work was supported by the National Natural Science Foundation of China under Grant Nos. 12071110 and 62271189, the Hebei Central Guidance on Local Science and Technology Development Foundation of China under Grant No. 236Z7604G, and the Doctoral Science Start Foundation of Hebei GEO University of China under Grant No. BQ2024075.
\end{acknowledgments}

\begin{appendix}
\section{The Theorem 1 of Ref. \cite{ZGY}}\label{A}
The \cite[Theorem 1]{ZGY} presents a sufficient condition to demonstrate the local irreducibility of a kind of OPSs. This class of OPSs is described by \cite{ZGY}
\begin{equation}\label{S}
\begin{aligned}
S=\cup_{r\in Q}S_{r},~~Q=\{1,2, \ldots,q\},
\end{aligned}
\end{equation}
where $S_{r}$ expresses the orthogonal product basis spanned by computational basis $\mathcal{B}_{r}=\mathcal{B}_{r}^{\{1\}}\otimes \mathcal{B}_{r}^{\{2\}}\otimes \cdots\otimes \mathcal{B}_{r}^{\{n\}}$ of the corresponding subspace and every component of the quantum state in $S_{r}$ is nonzero under this computational basis. It should be noted that $\mathcal{B}_{r}$ ($1\leq r \leq q$) are disjoint and $\mathcal{B}_{r}^{\{k\}}\subset \mathcal{B}^{\{k\}}$ for every $k$. Here are some necessary notations and concepts \cite{ZGY}.

Assume $X=\{k_{1},\ldots,k_{m}\}$ is a subset of $\{1,2,\ldots,n\}$, $\overline{X}$ is the complement of $X$. For the subset $S_{r}$ in (\ref{S}), there are concepts.

(c1) Projection set on the $X$ party: $\mathcal{S}_{r}^{(X)}=\mathcal{B}_{r}^{\{k_{1}\}}\otimes\cdots\otimes\mathcal{B}_{r}^{\{k_{m}\}}$.

(c2) Projection inclusion (PI) set on party $X$: $R_{r}=\cup_{t}\mathcal{S}_{t}$ $(t\notin r)$ satisfying  $S_{r}^{(X)}\subset \bigcup_{t}S_{t}^{(X)}$ and $\bigcap_{t}S_{t}^{(\overline{X})}\neq \emptyset$.

(c3) More useful projection inclusion (UPI) set: $R_{r}$, where there exists a subset $S_{t}\subset R_{r}$ such that $\big|S_{r}^{(X)}\bigcap S_{t}^{(X)}\big|=1$.

For the set $S$ in (\ref{S}), we review a set sequence $G_{1},G_{2},\ldots,G_{s}$, where $G_{x}=\cup_{r_{x}}S_{r_{x}}$ $(x=1,\ldots,s)$. It satisfies the following three conditions.

(g1) The sets $G_{x}$ are pairwise disjoint and $\cup_{x=1}^{s}G_{x}=S$.

(g2) $G_{1}$ is the union of all subsets $S_{r_{1}}$ that have UPI sets.

(g3) For any $S_{r_{x+1}}\subset G_{x+1}$ $(x=1,\ldots,s-1)$, there is at least one $S_{r_{x}}\subset G_{x}$ such that $S_{r_{x}}^{(X)}\cap S_{r_{x+1}}^{(X)}\neq\emptyset$.

Note that such a set sequence $G_{1},G_{2},\ldots,G_{s}$ satisfying above (g1)-(g3) does not necessarily exist.

In addition, the family of projection sets $\{S_{r}^{(X)}\}_{r\in Q}$ is called connected if it cannot be divided into two groups of sets $\{S_{k}^{(X)}\}_{k\in T}$ $(T\subsetneqq Q)$ and $\{S_{l}^{(X)}\}_{l\in Q\setminus T}$ such that
\begin{equation}\label{1}
\begin{aligned}
\bigg(\bigcup\limits_{k\in T}S_{k}^{(X)}\bigg)\bigcap \bigg(\bigcup\limits_{l\in Q\setminus T}S_{l}^{(X)}\bigg)=\emptyset.
\end{aligned}
\end{equation}

Let $\mathcal{B}^{X}=\{|i\rangle_{X}\}_{i=0}^{d_{X}-1}=\{\otimes_{j=1}^{m} |i_{k_{j}}\rangle|i_{k_{j}}=0,\ldots,d_{k_{j}}-1\}$ be the computational basis of party $X$, where $d_{X}=d_{k_{1}}\cdots d_{k_{m}}$.
For a fixed $i\in\mathcal{Z}_{d_{X}}$, $\mathcal{B}_{i}^{X}:=\{|l\rangle_{X}\}_{l=i}^{d_{X}-1}$, $V_{i}:=\{\bigcup_{v}S_{v}^{(\overline{X})}~|~|i\rangle_{X}\in S_{v}^{(X)}\}$, and $\widetilde{S}_{V_{i}}:=\{\bigcup_{s}S_{s}^{(X)}~|~S_{s}^{(\overline{X})}\cap V_{i}\neq \emptyset\}$. There is the \cite[Theorem 1]{ZGY}.

\cite[Theorem 1]{ZGY}. For the given set $S$ in (\ref{S}), any orthogonality-preserving POVM performed on party $X$ can only be trivial if the following conditions are satisfied.

(i) There is a relationship $\mathcal{B}_{i}^{X}\subset \widetilde{S}_{V_{i}}$ for any $i\in\mathcal{Z}_{d_{X}-1}$.

(ii) For every $S_{r}$, there exists a corresponding PI set $R_{r}$ on $X$ party.

(iii) The set sequence $G_{1},\ldots,G_{s}$ satisfying (g1)-(g3) exists. Moreover, for each $S_{r_{x+1}}\subset G_{x+1}$ with $x=1,2,\ldots,s-1$, there is a $S_{r_{x}}\subset G_{x}$ and a $S_{r_{x+1}'}\subset R_{r_{x+1}}$ such that $S_{r_{x}}^{(X)}\cap S_{r_{x+1}}^{(X)}\supset S_{r_{x+1}}^{(X)}\cap S_{r_{x+1}'}^{(X)}$.

(iv) The family of sets $\{S_{r}^{(X)}\}_{r\in Q}$ is connected.

\section{The proof of theorem 2}\label{B}
There are six situations for the union of any two subsystems, i.e., $X_{12}$, $X_{13}$, $X_{14}$, $X_{23}$, $X_{24}$, and $X_{34}$. Combining the Theorem 1 and \cite[Theorem 1]{ZGY}, we only need to confirm that there exists a subset that satisfies the four conditions of \cite[Theorem 1]{ZGY} on every union.

\begin{figure}[h]
\centering
\includegraphics[width=0.41\textwidth]{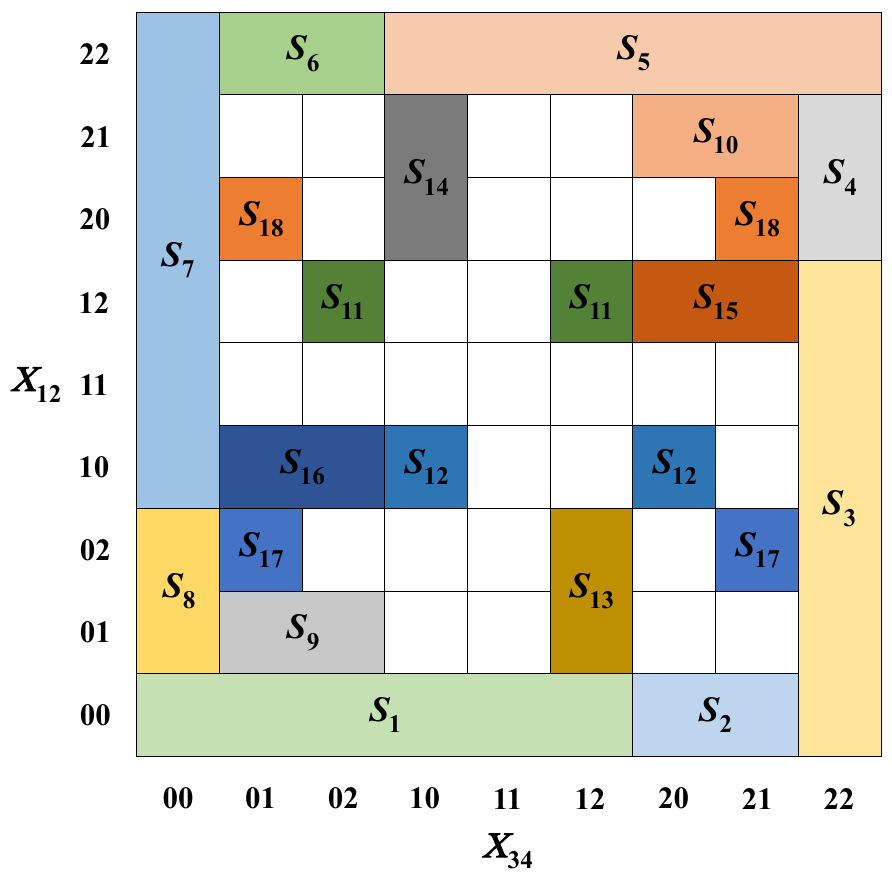}
\caption{The corresponding $9\times 9$ grid of $\{S_{r}\}_{r=1}^{18}$ given by Eq. (\ref{21}) in $X_{12}|X_{34}$ bipartition. \label{B1}}
\end{figure}

In $X_{12}|X_{34}$ bipartition, we consider the subset $\{S_{r}\}_{r=1}^{8}$ of $\{S_{r}\}_{r=1}^{18}$ given by Eq. (\ref{21}). The plane structure of $\{S_{r}\}_{r=1}^{18}$ (\ref{21}) is shown in Fig. \ref{B1}.

On party $X_{34}$, we find

(i) $\widetilde{S}_{V_{i}}=\mathcal{B}^{X_{34}}$, it is obvious that $\mathcal{B}_{i}^{X_{34}}\subset \widetilde{S}_{V_{i}}$ for any $i\in\mathcal{Z}_{8}$.

(ii) For every subset of $\{S_{r}\}_{r=1}^{8}$, there is the corresponding PI set. Specifically, $R_{1}=S_{5}\cup S_{6}\cup S_{7}$, $R_{2}=S_{5}$, $R_{3}=S_{5}$, $R_{4}=S_{5}$, $R_{5}=S_{1}\cup S_{2}\cup S_{3}$, $R_{6}=S_{1}$, $R_{7}=S_{1}$, and $R_{8}=S_{1}$.

(iii) The set sequence
\begin{equation}
\begin{aligned}
&G_{1}=S_{1}\cup S_{3}\cup S_{4}\cup S_{5}\cup S_{7}\cup S_{8},\\
&G_{2}=S_{2}\cup S_{6},
\end{aligned}
\end{equation}
satisfies (g1)-(g3). Moreover, for $S_{2}\subset G_{2}$ and $S_{6}\subset G_{2}$, $S_{5}=G_{1}\cap R_{2}$ and $S_{1}=G_{1}\cap R_{6}$ meet the requirement, respectively.

(iv) We find $S_{1}^{(X_{34})}\cap S_{5}^{(X_{34})}\neq \emptyset$ and $S_{1}^{(X_{34})}\cup S_{5}^{(X_{34})}=\mathcal{B}^{(X_{34})}$, which means that the family of projection sets of $\{S_{r}\}_{r=1}^{8}$ is connected.

Thus, for the set $\{S_{r}\}_{r=1}^{8}$, the orthogonality-preserving POVM performed on party $X_{34}$ is only trivial. On party $X_{12}$, it is similar.

\begin{figure}[h]
\centering
\includegraphics[width=0.4\textwidth]{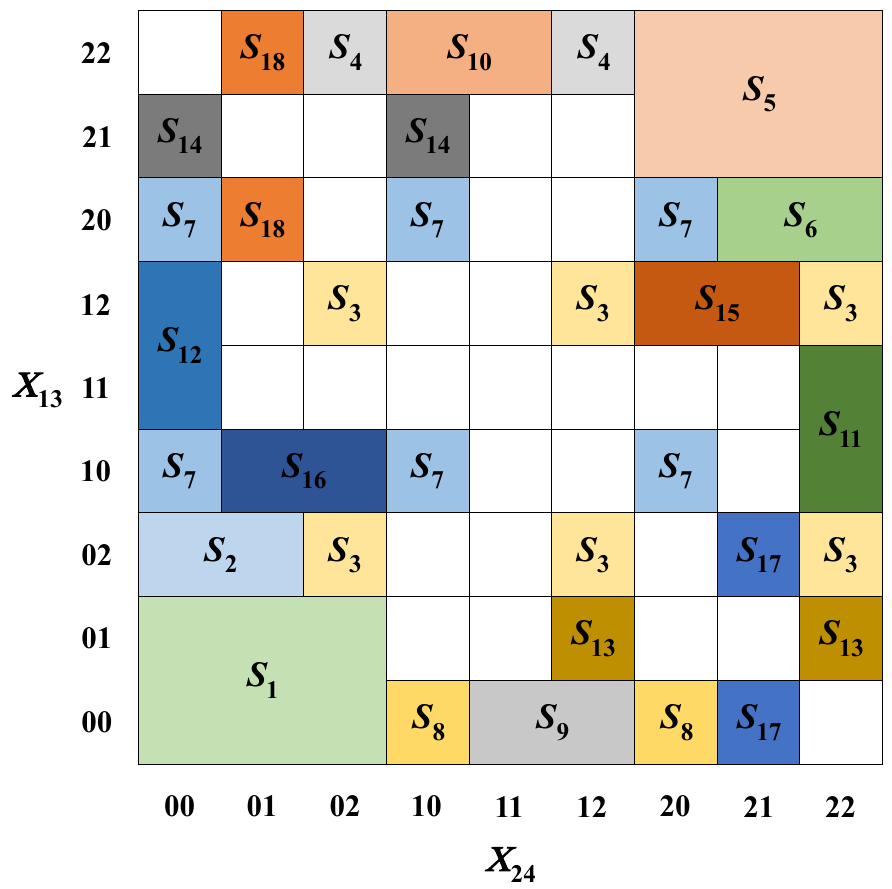}
\caption{The corresponding $9\times 9$ grid of $\{S_{r}\}_{r=1}^{18}$ given by Eq. (\ref{21}) in $X_{13}|X_{24}$ bipartition. \label{B2}}
\end{figure}

In $X_{13}|X_{24}$ bipartition, the plane structure of $\{S_{r}\}_{r=1}^{18}$ (\ref{21}) is shown in Fig. \ref{B2}.

On party $X_{24}$, we can know

(i) $\mathcal{B}_{i}^{X_{24}}\subset\widetilde{S}_{V_{i}}=\mathcal{B}^{X_{24}}$ for any $i\in\mathcal{Z}_{8}$.

(ii) For each subset $S_{r}$, the corresponding PI set $R_{r}$ is given by the following Table \ref{Bt1}.
\begin{table}[tbp]
\centering
\caption{Corresponding PI set $R_{r}$ for each subset $S_{r}$ on party $X_{24}$.}\label{Bt1}
\begin{tabular}{cl|cl}
\hline
\hline
Subset~~ & ~~~~~PI set~~~~~ & ~~Subset~~ & ~~~~~PI set~~~~~ \\ \hline
 $S_{1}$ & $R_{1}=S_{2}\cup S_{3}$ & $S_{10}$ & $R_{10}=S_{8}\cup S_{9}$ \\
 $S_{2}$ & $R_{2}=S_{7}\cup S_{16}$ & $S_{11}$ & $R_{11}=S_{5}$ \\
 $S_{3}$ & $R_{3}=S_{4}\cup S_{5}$  & $S_{12}$ & $R_{12}=S_{1}$  \\
 $S_{4}$ & $R_{4}=S_{1}\cup S_{9}$ & $S_{13}$ & $R_{13}=S_{4}\cup S_{5}$ \\
 $S_{5}$ & $R_{5}=S_{6}\cup S_{7}$ & $S_{14}$ & $R_{14}=S_{1}\cup S_{8}$ \\
 $S_{6}$ & $R_{6}=S_{3}\cup S_{15}$ & $S_{15}$ & $R_{15}=S_{6}\cup S_{7}$  \\
 $S_{7}$ & $R_{7}=S_{1}\cup S_{8}$ & $S_{16}$ & $R_{16}=S_{2}\cup S_{3}$ \\
 $S_{8}$ & $R_{8}=S_{5}\cup S_{10}$ & $S_{17}$ & $R_{17}=S_{5}$ \\
 $S_{9}$ & $R_{9}=S_{4}\cup S_{10}$ & $S_{18}$ & $R_{18}=S_{1}$ \\
\hline
\end{tabular}
\end{table}

(iii) There is the set sequence $G_{1}=\{S_{r}\}_{r=1}^{18}$. Obviously, it holds.

(iv) For the sequence of projection sets $S_{1}^{(X_{24})}\rightarrow S_{4}^{(X_{24})}\rightarrow S_{9}^{(X_{24})}\rightarrow S_{10}^{(X_{24})}\rightarrow S_{8}^{(X_{24})}\rightarrow S_{5}^{(X_{24})}$, the intersection of two adjacent sets is not empty and their union is the computation basis $\mathcal{B}^{X_{24}}$. This means that the family of projection sets $\{S_{r}^{(X_{24})}\}_{r}$ is connected.

So, the orthogonality-preserving POVM performed on party $X_{24}$ can only be trivial. There is a similar situation on the $X_{13}$ party.

\begin{figure}[h]
\centering
\includegraphics[width=0.4\textwidth]{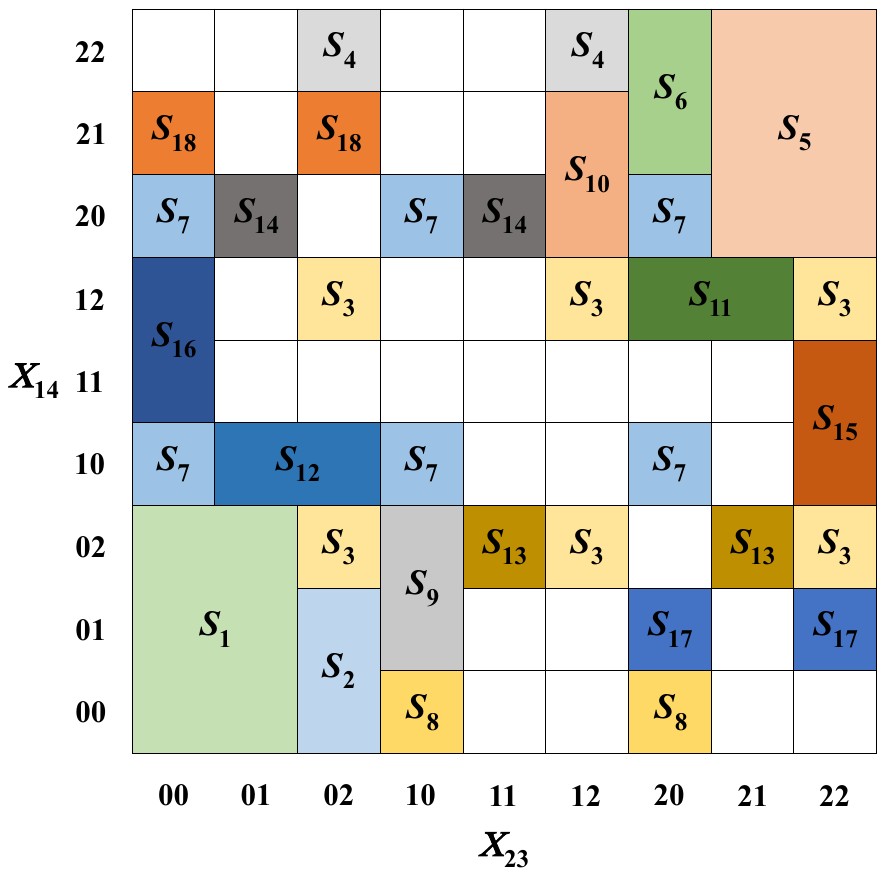}
\caption{The corresponding $9\times 9$ grid of $\{S_{r}\}_{r=1}^{18}$ given by Eq. (\ref{21}) in $X_{14}|X_{23}$ bipartition. \label{B3}}
\end{figure}

In $X_{14}|X_{23}$ bipartition, a geometric representation of $\{S_{r}\}_{r=1}^{18}$ (\ref{21}) is given by Fig. \ref{B3}.

Considering $X_{23}$ party, we have

(i) $\mathcal{B}_{i}^{X_{23}}\subset\widetilde{S}_{V_{i}}=\mathcal{B}^{X_{23}}$ for any $i\in\mathcal{Z}_{8}$.

(ii) For every $S_{r}$, the corresponding PI set $R_{r}$ is shown in Table \ref{Bt2}.
\begin{table}[tbp]
\centering
\caption{Corresponding PI set $R_{r}$ for each subset $S_{r}$ on the $X_{23}$ party.}\label{Bt2}
\begin{tabular}{cl|cl}
\hline
\hline
Subset~~ & ~~~~~PI set~~~~~ & ~~Subset~~ & ~~~~~PI set~~~~~ \\ \hline
 $S_{1}$ & $R_{1}=S_{7}\cup S_{12}$  & $S_{10}$ & $R_{10}=S_{3}$ \\
 $S_{2}$ & $R_{2}=S_{3}$             & $S_{11}$ & $R_{11}=S_{5}\cup S_{6}$ \\
 $S_{3}$ & $R_{3}=S_{4}\cup S_{5}$   & $S_{12}$ & $R_{12}=S_{1}\cup S_{2}$  \\
 $S_{4}$ & $R_{4}=S_{10}\cup S_{18}$ & $S_{13}$ & $R_{13}=S_{5}\cup S_{14}$ \\
 $S_{5}$ & $R_{5}=S_{3}\cup S_{11}$  & $S_{14}$ & $R_{14}=S_{1}\cup S_{13}$ \\
 $S_{6}$ & $R_{6}=S_{7}$             & $S_{15}$ & $R_{15}=S_{3}$  \\
 $S_{7}$ & $R_{7}=S_{1}\cup S_{8}$   & $S_{16}$ & $R_{16}=S_{7}$ \\
 $S_{8}$ & $R_{8}=S_{9}\cup S_{17}$  & $S_{17}$ & $R_{17}=S_{5}\cup S_{6}$ \\
 $S_{9}$ & $R_{9}=S_{7}$             & $S_{18}$ & $R_{18}=S_{1}\cup S_{2}$ \\
\hline
\end{tabular}
\end{table}

(iii) The set sequence $G_{1}=\{S_{r}\}_{r=1}^{18}$ holds.

(iv) There is a sequence of projection sets $S_{1}^{(X_{23})}\rightarrow S_{12}^{(X_{23})}\rightarrow S_{3}^{(X_{23})}\rightarrow S_{5}^{(X_{23})}\rightarrow S_{11}^{(X_{23})}(\rightarrow S_{13}^{(X_{23})})\rightarrow S_{7}^{(X_{23})}$, where the set $S_{13}^{(X_{23})}$ in parentheses is only related to the previous set $S_{11}^{(X_{23})}$. It satisfies that the intersection of two adjacent sets is not empty and their union is the computation basis $\mathcal{B}^{X_{23}}$. This means that the family of projection sets $\{S_{r}^{(X_{23})}\}_{r}$ is connected.

Hence, the orthogonality-preserving POVM performed on party $X_{23}$ can only be trivial. The same goes for the $X_{14}$ party.

To sum up, the set $\{S_{r}\}_{r=1}^{18}$ (\ref{21}) is strongly nonlocal in every three-partition.

\section{The proof of theorem 3}\label{C}
It is similar to the proof of Theorem 2. We discuss the orthogonality-preserving POVM performed on parties $X_{12}$, $X_{13}$, $X_{14}$, $X_{23}$, $X_{24}$, and $X_{34}$, respectively. Primarily, we need to prove that the set $\{H_{r}\}_{r=1}^{18}$ (\ref{22}) satisfies the four conditions in the \cite[Theorem 1]{ZGY} on each party. Obviously, the conditions (i), (ii), and (iv) still hold, since the OPS (\ref{22}) has the same structure as the OPS (\ref{21}). Next, we will illustrate the condition (iii).

\begin{figure}[h]
\centering
\includegraphics[width=0.41\textwidth]{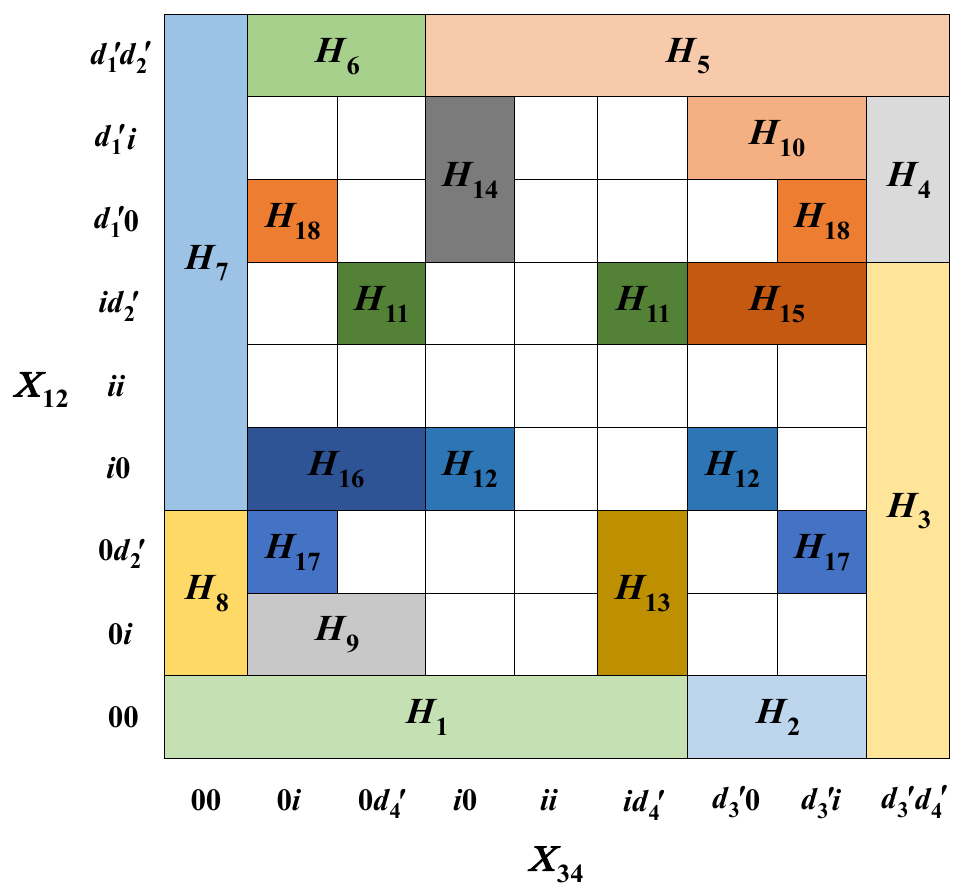}
\caption{The corresponding $9\times 9$ grid of $\{H_{r}\}_{r=1}^{18}$ given by Eq. (\ref{22}) in $X_{12}|X_{34}$ bipartition. Here $i$ of the $k$th subsystem expresses $\{1,\ldots,d_{k}'-1\}$ with $k=1,2,3,4$. \label{B4}}
\end{figure}

On party $X_{34}$, we can refer to the Fig. \ref{B4}, which is a geometric representation of $\{H_{r}\}_{r=1}^{18}$ (\ref{22}) in $X_{12}|X_{34}$ bipartition. Consider the subset $\{H_{r}\}_{r=1}^{8}$ of OPS (\ref{22}), there is the set sequence
\begin{equation}
\begin{aligned}
&G_{1}=H_{1}\cup H_{3}\cup H_{4}\cup H_{5}\cup H_{7}\cup H_{8},\\
&G_{2}=H_{2}\cup H_{6}.
\end{aligned}
\end{equation}
For the subsets $H_{2}\subset G_{2}$ and $H_{6}\subset G_{2}$, there are the PI sets $R_{2}=H_{5}$ and $R_{6}=H_{1}$ on party $X_{34}$, respectively. It is easy to conclude that $H_{5}=G_{1}\cap R_{2}$ and $H_{1}=G_{1}\cap R_{6}$. This means that the condition (iii) is true.

Similarly, on party $X_{24}$, the set sequence has
\begin{equation}
\begin{aligned}
&G_{1}=H_{1}\cup H_{2}\cup H_{3}\cup H_{4}\cup H_{5}\cup H_{6}\cup H_{7}\cup H_{8}\\
&\quad~~~~~~\cup H_{11}\cup H_{12}\cup H_{13}\cup H_{14}\cup H_{15}\cup H_{16},\\
&G_{2}=H_{9}\cup H_{10}.
\end{aligned}
\end{equation}
For the subsets $H_{9}\subset G_{2}$ and $H_{10}\subset G_{2}$, there are the PI sets $R_{9}=H_{4}\cup H_{10}$ and $R_{10}=H_{8}\cup H_{9}$ on party $X_{24}$, respectively. Then, we have $H_{4}=G_{1}\cap R_{9}$ and $H_{8}=G_{1}\cap R_{10}$.

On party $X_{23}$, the set sequence has
\begin{equation}
\begin{aligned}
&G_{1}=H_{1}\cup H_{2}\cup H_{3}\cup H_{4}\cup H_{5}\cup H_{6}\cup H_{7}\\
&\quad~~~~~~\cup H_{8}\cup H_{11}\cup H_{12}\cup H_{15}\cup H_{16},\\
&G_{2}=H_{9}\cup H_{10}\cup H_{13}\cup H_{14}.
\end{aligned}
\end{equation}
Referring to Table \ref{Bt2}, for the subsets $H_{9}\subset G_{2}$, $H_{10}\subset G_{2}$, $H_{13}\subset G_{2}$, and $H_{14}\subset G_{2}$, there are PI sets $R_{9}=H_{7}$, $R_{10}=H_{3}$, $R_{13}=H_{5}\cup H_{14}$, and $R_{14}=H_{1}\cup H_{13}$ on party $X_{23}$, respectively. Naturally, we get $H_{7}=G_{1}\cap R_{9}$, $H_{3}=G_{1}\cap R_{10}$, $H_{5}=G_{1}\cap R_{13}$, and $H_{1}=G_{1}\cap R_{14}$.

For the remaining three parties $X_{12}$, $X_{13}$, and $X_{14}$, there are the similar results. So, the orthogonality-preserving POVM performed on each party can only be trivial. The set $\{H_{r}\}_{r=1}^{18}$ (\ref{22}) is strongly nonlocal in every three-partition.

\section{The proof of theorem 4}\label{D}
\begin{figure}[h]
\centering
\includegraphics[width=0.4\textwidth]{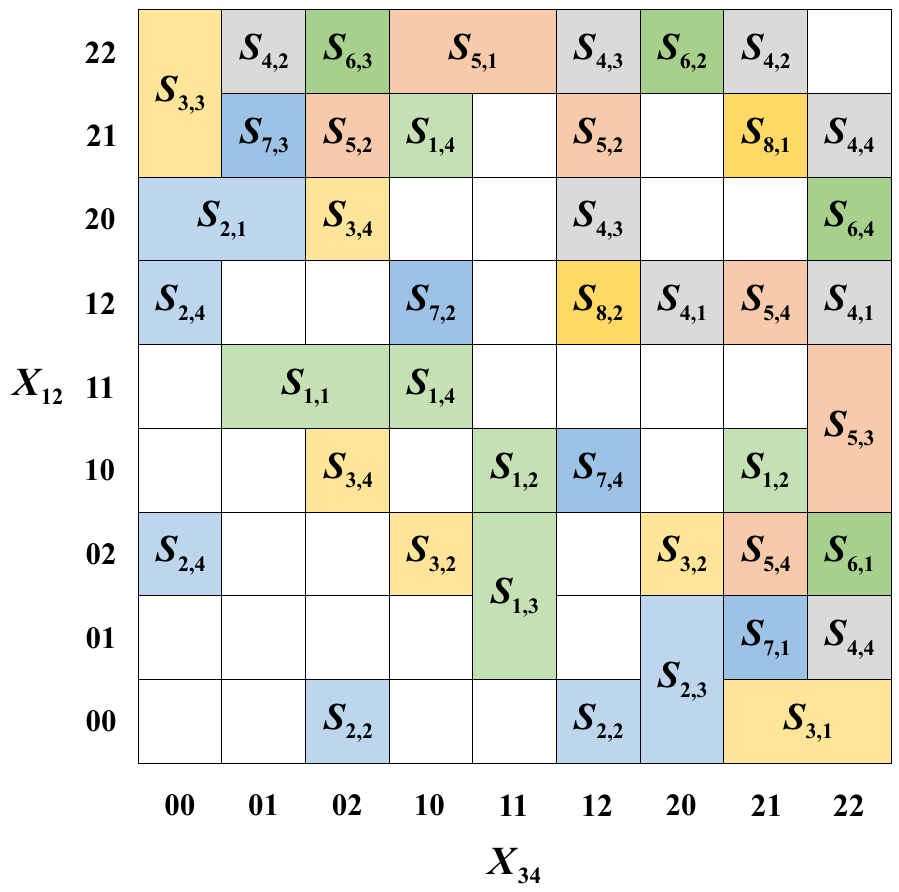}
\caption{The corresponding $9\times 9$ grid of $\{S_{r,t}\}$ given by Eq. (\ref{23}) in $X_{12}|X_{34}$ bipartition. \label{B5}}
\end{figure}

Due to the symmetry, we only need to discuss two situations, parties $X_{34}$ and $X_{24}$. In $X_{12}|X_{34}$ bipartition, a geometric representation of $\{S_{r,t}\}$ (\ref{23}) is given by Fig. \ref{B5}.

On party $X_{34}$, there are

(i) $\mathcal{B}_{i}^{X_{34}}\subset\widetilde{S}_{V_{i}}=\mathcal{B}^{X_{34}}$ for any $i\in\mathcal{Z}_{8}$.

(ii) For every $S_{r,t}$, the corresponding PI set $R_{r,t}$ is shown in Table \ref{Bt3}.
\begin{table}[tbp]
\centering
\caption{Corresponding PI set $R_{r,t}$ for each subset $S_{r,t}$ on the $X_{34}$ party.}\label{Bt3}
\begin{tabular}{cl|cl}
\hline
\hline
Subset~~ & ~~~~~PI set~~~~~ & ~~Subset~~ & ~~~~~PI set~~~~~ \\ \hline
 $S_{1,1}$ & $R_{1,1}=S_{4,2}\cup S_{6,3}$ & $S_{4,4}$ & $R_{4,4}=S_{3,1}$ \\
 $S_{1,2}$ & $R_{1,2}=S_{5,1}\cup S_{4,2}$ & $S_{5,1}$ & $R_{5,1}=S_{1,3}\cup S_{3,2}$ \\
 $S_{1,3}$ & $R_{1,3}=S_{5,1}$             & $S_{5,2}$ & $R_{5,2}=S_{4,3}\cup S_{6,3}$  \\
 $S_{1,4}$ & $R_{1,4}=S_{3,2}$             & $S_{5,3}$ & $R_{5,3}=S_{3,1}$ \\
 $S_{2,1}$ & $R_{2,1}=S_{3,3}\cup S_{4,2}$ & $S_{5,4}$ & $R_{5,4}=S_{3,1}$ \\
 $S_{2,2}$ & $R_{2,2}=S_{6,3}\cup S_{4,3}$ & $S_{6,1}$ & $R_{6,1}=S_{3,1}$ \\
 $S_{2,3}$ & $R_{2,3}=S_{6,2}$             & $S_{6,2}$ & $R_{6,2}=S_{2,3}$ \\
 $S_{2,4}$ & $R_{2,4}=S_{3,3}$             & $S_{6,3}$ & $R_{6,3}=S_{5,2}$ \\
 $S_{3,1}$ & $R_{3,1}=S_{8,1}\cup S_{4,4}$ & $S_{6,4}$ & $R_{6,4}=S_{3,1}$ \\
 $S_{3,2}$ & $R_{3,2}=S_{5,1}\cup S_{6,2}$ & $S_{7,1}$ & $R_{7,1}=S_{3,1}$ \\
 $S_{3,3}$ & $R_{3,3}=S_{2,1}$             & $S_{7,2}$ & $R_{7,2}=S_{3,2}$ \\
 $S_{3,4}$ & $R_{3,4}=S_{6,3}$             & $S_{7,3}$ & $R_{7,3}=S_{2,1}$ \\
 $S_{4,1}$ & $R_{4,1}=S_{2,3}\cup S_{3,1}$ & $S_{7,4}$ & $R_{7,4}=S_{2,2}$ \\
 $S_{4,2}$ & $R_{4,2}=S_{7,3}\cup S_{8,1}$ & $S_{8,1}$ & $R_{8,1}=S_{3,1}$ \\
 $S_{4,3}$ & $R_{4,3}=S_{5,2}$             & $S_{8,2}$ & $R_{8,2}=S_{5,2}$ \\
\hline
\end{tabular}
\end{table}

(iii) The set sequence $G_{1}=\cup_{r,t}S_{r,t}$ holds.

(iv) There is a sequence of projection sets $S_{5,1}^{(X_{34})}\rightarrow S_{3,2}^{(X_{34})}\rightarrow S_{4,1}^{(X_{34})}\rightarrow S_{3,1}^{(X_{34})}\rightarrow S_{4,2}^{(X_{34})}\rightarrow S_{2,1}^{(X_{34})}\rightarrow S_{1,1}^{(X_{34})}\rightarrow S_{5,2}^{(X_{34})}$. Here, the intersection of two adjacent sets is not empty and their union is the computation basis $\mathcal{B}^{X_{34}}$. So, the family of projection sets $\{S_{r,t}^{(X_{34})}\}$ is connected.

\begin{figure}[h]
\centering
\includegraphics[width=0.4\textwidth]{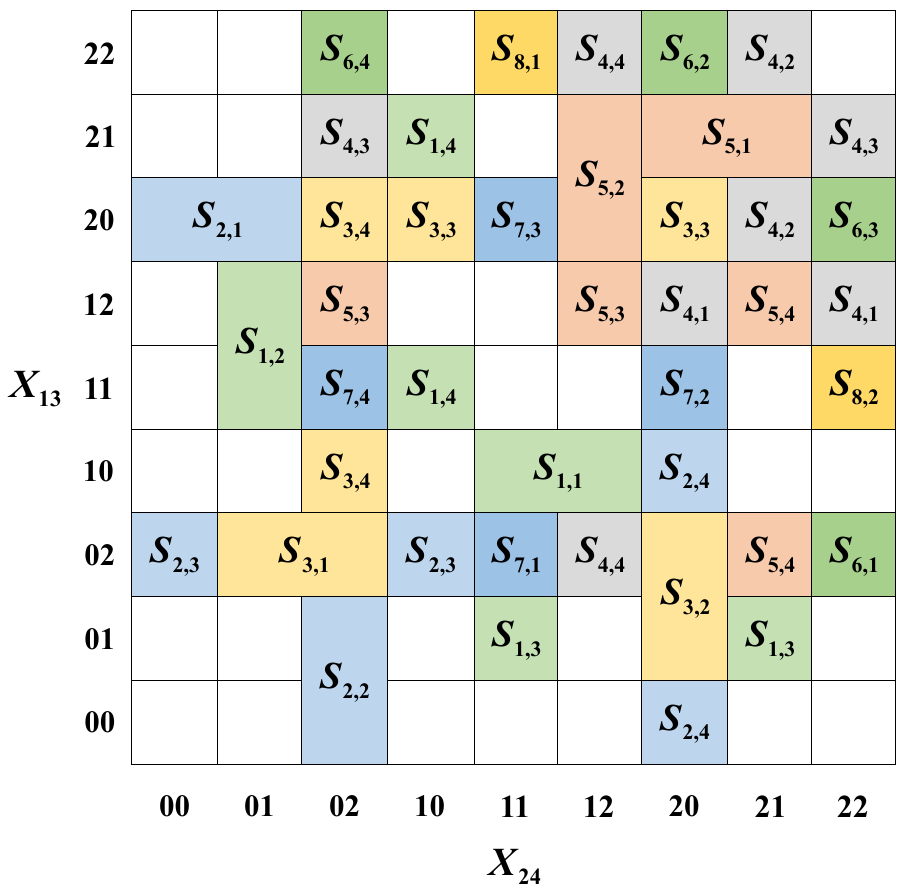}
\caption{The corresponding $9\times 9$ grid of $\{S_{r,t}\}$ given by Eq. (\ref{23}) in $X_{13}|X_{24}$ bipartition. \label{B6}}
\end{figure}

Fig. \ref{B6} shows the planar structure of $\{S_{r,t}\}$ (\ref{23}) in $X_{13}|X_{24}$ bipartition. On party $X_{24}$, we have

(i) $\mathcal{B}_{i}^{X_{24}}\subset\widetilde{S}_{V_{i}}=\mathcal{B}^{X_{24}}$ for any $i\in\mathcal{Z}_{8}$.

(ii) For every $S_{r,t}$, the corresponding PI set $R_{r,t}$ is shown in Table \ref{Bt4}.
\begin{table}[tbp]
\centering
\caption{Corresponding PI set $R_{r,t}$ for each subset $S_{r,t}$ on the $X_{24}$ party.}\label{Bt4}
\begin{tabular}{cl|cl}
\hline
\hline
Subset~~ & ~~~~~PI set~~~~~ & ~~Subset~~ & ~~~~~PI set~~~~~ \\ \hline
 $S_{1,1}$ & $R_{1,1}=S_{7,3}\cup S_{5,2}$ & $S_{4,4}$ & $R_{4,4}=S_{5,3}$ \\
 $S_{1,2}$ & $R_{1,2}=S_{2,1}$             & $S_{5,1}$ & $R_{5,1}=S_{3,3}\cup S_{4,2}$ \\
 $S_{1,3}$ & $R_{1,3}=S_{5,4}\cup S_{7,1}$ & $S_{5,2}$ & $R_{5,2}=S_{5,3}$  \\
 $S_{1,4}$ & $R_{1,4}=S_{3,3}$             & $S_{5,3}$ & $R_{5,3}=S_{3,1}\cup S_{4,4}$ \\
 $S_{2,1}$ & $R_{2,1}=S_{2,3}\cup S_{3,1}$ & $S_{5,4}$ & $R_{5,4}=S_{5,1}$ \\
 $S_{2,2}$ & $R_{2,2}=S_{3,1}$             & $S_{6,1}$ & $R_{6,1}=S_{4,1}$ \\
 $S_{2,3}$ & $R_{2,3}=S_{2,1}\cup S_{3,3}$ & $S_{6,2}$ & $R_{6,2}=S_{2,4}$ \\
 $S_{2,4}$ & $R_{2,4}=S_{5,1}$             & $S_{6,3}$ & $R_{6,3}=S_{8,2}$ \\
 $S_{3,1}$ & $R_{3,1}=S_{2,1}\cup S_{3,4}$ & $S_{6,4}$ & $R_{6,4}=S_{3,1}$ \\
 $S_{3,2}$ & $R_{3,2}=S_{5,1}$             & $S_{7,1}$ & $R_{7,1}=S_{1,1}$ \\
 $S_{3,3}$ & $R_{3,3}=S_{1,4}\cup S_{5,1}$ & $S_{7,2}$ & $R_{7,2}=S_{3,3}$ \\
 $S_{3,4}$ & $R_{3,4}=S_{3,1}$             & $S_{7,3}$ & $R_{7,3}=S_{1,1}$ \\
 $S_{4,1}$ & $R_{4,1}=S_{4,3}\cup S_{5,1}$ & $S_{7,4}$ & $R_{7,4}=S_{3,1}$ \\
 $S_{4,2}$ & $R_{4,2}=S_{5,1}$             & $S_{8,1}$ & $R_{8,1}=S_{1,1}$ \\
 $S_{4,3}$ & $R_{4,3}=S_{3,1}\cup S_{6,1}$ & $S_{8,2}$ & $R_{8,2}=S_{4,1}$ \\
\hline
\end{tabular}
\end{table}

(iii) The set sequence $G_{1}=\cup_{r,t}S_{r,t}$ holds.

(iv) There is a sequence of projection sets $S_{2,3}^{(X_{24})}\rightarrow S_{2,1}^{(X_{24})}\rightarrow S_{3,1}^{(X_{24})}\rightarrow S_{5,3}^{(X_{24})}\rightarrow S_{1,1}^{(X_{24})}\rightarrow S_{1,3}^{(X_{24})}\rightarrow S_{5,1}^{(X_{24})}\rightarrow S_{4,1}^{(X_{24})}$. It satisfies that the intersection of two adjacent sets is not empty and their union is the computation basis $\mathcal{B}^{X_{24}}$. This means that the family of projection sets $\{S_{r,t}^{(X_{24})}\}$ is connected.

On parties $X_{34}$ and $X_{24}$, the OPS of Theorem 4 satisfies the four conditions in the \cite[Theorem 1]{ZGY}. Therefore, this OPS has the strong nonlocality in every three-partition.

\section{The proof of theorem 5}\label{E}
Due to the symmetry, we only need to consider the four conditions of \cite[Theorem 1]{ZGY} on parties $X_{34}$ and $X_{24}$, respectively. The OPS $\cup_{r,t}H_{r,t}$ (\ref{24}) in Theorem 5 is the generalization of OPS $\cup_{r,t}S_{r,t}$ (\ref{23}) in Theorem 4. They have the same structure. Obviously, the conditions (i), (ii), and (iv) still hold. For the remaining condition (iii), we have the following results.

On party $X_{34}$, there is the set sequence
\begin{equation}
\begin{aligned}
&G_{1}=H_{1,1}\cup H_{2,1}\cup H_{2,2}\cup H_{2,3}\cup H_{2,4}\cup H_{3,1}\\
&\quad~~~~~~\cup H_{3,2}\cup H_{3,3}\cup H_{3,4}\cup H_{4,1}\cup H_{4,4}\\
&\quad~~~~~~\cup H_{5,2}\cup H_{5,3}\cup H_{6,1}\cup H_{6,2}\cup H_{6,3}\\
&\quad~~~~~~\cup H_{6,4},\\
&G_{2}=H_{1,4}\cup H_{4,3}\cup H_{5,1}\cup H_{5,4}\cup H_{7,1}\cup H_{7,2}\\
&\quad~~~~~~\cup H_{7,3}\cup H_{7,4}\cup H_{8,1}\cup H_{8,2},\\
&G_{3}=H_{1,2}\cup H_{1,3}\cup H_{4,2}.
\end{aligned}
\end{equation}
The PI set $R_{r,t}$ of each subset $H_{r,t}$ corresponds to the Table \ref{Bt3}. Then, for every subset $H_{r,t}\subset G_{x}$ $(x=2,3)$, we get the intersection of set $G_{x-1}$ and PI set $R_{r,t}$, which is shown in Table \ref{Bt5}.
\begin{table}[tbp]
\centering
\caption{The intersection of set $G_{x-1}$ and PI set $R_{r,t}$ about subset $H_{r,t}\subset G_{x}$ $(x=2,3)$.}\label{Bt5}
\begin{tabular}{cc|cc}
\hline
\hline
~~~Subset~~~ & ~~~~~Intersection~~~~~ & ~~~Subset~~~ & ~~~~~Intersection~~~~~   \\ \hline
 $H_{1,4}\subset G_{2}$ & $H_{3,2}=G_{1}\cap R_{1,4}$ & $H_{7,4}\subset G_{2}$ & $H_{2,2}=G_{1}\cap R_{7,4}$ \\
 $H_{4,3}\subset G_{2}$ & $H_{5,2}=G_{1}\cap R_{4,3}$ & $H_{8,1}\subset G_{2}$ & $H_{3,1}=G_{1}\cap R_{8,1}$ \\
 $H_{5,1}\subset G_{2}$ & $H_{3,2}=G_{1}\cap R_{5,1}$ & $H_{8,2}\subset G_{2}$ & $H_{5,2}=G_{1}\cap R_{8,2}$ \\
 $H_{5,4}\subset G_{2}$ & $H_{3,1}=G_{1}\cap R_{5,4}$ & $H_{1,2}\subset G_{3}$ & $H_{5,1}=G_{2}\cap R_{1,2}$ \\
 $H_{7,1}\subset G_{2}$ & $H_{3,1}=G_{1}\cap R_{7,1}$ & $H_{1,3}\subset G_{3}$ & $H_{5,1}=G_{2}\cap R_{1,3}$ \\
 $H_{7,2}\subset G_{2}$ & $H_{3,2}=G_{1}\cap R_{7,2}$ & $H_{4,2}\subset G_{3}$ & $H_{7,3}\subset G_{2}\cap R_{4,2}$ \\
 $H_{7,3}\subset G_{2}$ & $H_{2,1}=G_{1}\cap R_{7,3}$ & \\
\hline
\end{tabular}
\end{table}

On party $X_{24}$, there is the set sequence
\begin{equation}
\begin{aligned}
&G_{1}=H_{2,1}\cup H_{2,2}\cup H_{2,3}\cup H_{2,4}\cup H_{3,1}\cup H_{3,2}\\
&\quad~~~~~~\cup H_{3,3}\cup H_{3,4}\cup H_{4,1}\cup H_{4,3}\cup H_{5,1}\\
&\quad~~~~~~\cup H_{5,3}\cup H_{6,1}\cup H_{6,2}\cup H_{6,3}\cup H_{6,4}\\
&\quad~~~~~~\cup H_{7,2}\cup H_{7,4}\cup H_{8,2},\\
&G_{2}=H_{1,2}\cup H_{1,4}\cup H_{4,2}\cup H_{4,4}\cup H_{5,2}\cup H_{5,4}\\
&G_{3}=H_{1,1}\cup H_{1,3}\\
&G_{4}=H_{7,1}\cup H_{7,3}\cup H_{8,1}.
\end{aligned}
\end{equation}
Similarly, for each subset $H_{r,t}\subset G_{x}$ $(x=2,3,4)$, we obtain the intersection of set $G_{x-1}$ and PI set $R_{r,t}$ as shown in Table \ref{Bt6}.
\begin{table}[tbp]
\centering
\caption{The intersection of set $G_{x-1}$ and PI set $R_{r,t}$ about subset $H_{r,t}\subset G_{x}$ $(x=2,3,4)$.}\label{Bt6}
\begin{tabular}{cc|cc}
\hline
\hline
~~~Subset~~~ & ~~~~~Intersection~~~~~ & ~~~Subset~~~ & ~~~~~Intersection~~~~~   \\ \hline
 $H_{1,2}\subset G_{2}$ & $H_{2,1}=G_{1}\cap R_{1,2}$ & $H_{1,1}\subset G_{3}$ & $H_{5,2}=G_{2}\cap R_{1,1}$ \\
 $H_{1,4}\subset G_{2}$ & $H_{3,3}=G_{1}\cap R_{1,4}$ & $H_{1,3}\subset G_{3}$ & $H_{5,4}=G_{2}\cap R_{1,3}$ \\
 $H_{4,2}\subset G_{2}$ & $H_{5,1}=G_{1}\cap R_{4,2}$ & $H_{7,1}\subset G_{4}$ & $H_{1,1}=G_{3}\cap R_{7,1}$ \\
 $H_{4,4}\subset G_{2}$ & $H_{5,3}=G_{1}\cap R_{4,4}$ & $H_{7,3}\subset G_{4}$ & $H_{1,1}=G_{3}\cap R_{7,3}$ \\
 $H_{5,2}\subset G_{2}$ & $H_{5,3}=G_{1}\cap R_{5,2}$ & $H_{8,1}\subset G_{4}$ & $H_{1,1}=G_{3}\cap R_{8,1}$ \\
 $H_{5,4}\subset G_{2}$ & $H_{5,1}=G_{1}\cap R_{5,4}$ & \\
\hline
\end{tabular}
\end{table}

To sum up, whether on party $X_{34}$ or party $X_{24}$, the OPS in Theorem 5 meets the condition (iii) of \cite[Theorem 1]{ZGY}. So, this OPS in Theorem 5 is strongly nonlocal in every three-partition.

\section{The proof of theorem 6}\label{F}
Similarly, we only need to consider the POVM on party $X_{1l}$ with $l=2,\ldots,\frac{n+1}{2}$.

On party $X_{1l}$ ($l=2,\ldots,\frac{n-1}{2}$), we choose the subsets $H_{11,1}^{O}$, $H_{p1,\frac{n+3-2l}{2}}^{O}$, $H_{p1,\frac{n+1}{2}}^{O}$, and $H_{11,\frac{n+3}{2}}^{O}$. It is worth noting that the indicators $p$ of these subsets are independent of each other. Let $\widehat{H}_{i,j}^{O}$ express the union $\bigcup_{p}H_{pi,j}^{O}$.

(i) Due to the subset $H_{11,1}^{O}$, it is obvious that $\mathcal{B}_{i}^{X_{1l}}\subset\widetilde{S}_{V_{i}}=\mathcal{B}^{X_{1l}}$ for any $i\in\mathcal{Z}_{d_{1}d_{l}}$.

(ii) $H_{11,1}^{O}$ is the PI set of the other subsets $H_{p1,\frac{n+3-2l}{2}}^{O}$, $H_{p1,\frac{n+1}{2}}^{O}$, and $H_{11,\frac{n+3}{2}}^{O}$. The union of these subsets $\widehat{H}_{1,\frac{n+3-2l}{2}}^{O}\bigcup \widehat{H}_{1,\frac{n+1}{2}}^{O}\bigcup H_{11,\frac{n+3}{2}}^{O}$ is the PI set of $H_{11,1}^{O}$.

(iii) We have the set sequence
\begin{equation}
\begin{aligned}
&G_{1}=H_{11,1}^{O}\bigcup \widehat{H}_{1,\frac{n+1}{2}}^{O}\bigcup H_{11,\frac{n+3}{2}}^{O},\\
&G_{2}=\widehat{H}_{1,\frac{n+3-2l}{2}}^{O}.
\end{aligned}
\end{equation}
For every subset $H_{p1,\frac{n+3-2l}{2}}^{O}\subset G_{2}$, there is $G_{1}\bigcap R_{p1,\frac{n+3-2l}{2}}^{O}=H_{11,1}^{O}$ because $R_{p1,\frac{n+3-2l}{2}}^{O}=H_{11,1}^{O}$. The third condition is true.

(iv) Because the projection set of $H_{11,1}^{O}$ on party $X_{1l}$ is the computation basis $\mathcal{B}^{X_{1l}}$, the family of projection sets $\big\{H_{11,1}^{O},H_{p1,\frac{n+3-2l}{2}}^{O},H_{p1,\frac{n+1}{2}}^{O},H_{11,\frac{n+3}{2}}^{O}\big\}^{(X_{1l})}$ is connected. Here and below $\{H_{pr,t}^{O}\}^{(X_{1l})}$ expresses the projection set of $H_{pr,t}^{O}$ on party $X_{1l}$.

On party $X_{1\frac{n+1}{2}}$, we choose the subsets $H_{p1,1}^{O}$, $H_{11,\frac{n-1}{2}}^{O}$, $H_{p1,\frac{n+1}{2}}^{O}$, $H_{11,\frac{n+3}{2}}^{O}$, and $H_{p2,1}^{O}$.

(i) Due to the subsets $H_{p1,1}^{O}$ and $H_{11,\frac{n-1}{2}}^{O}$, it is obvious that $\mathcal{B}_{i}^{X_{1\frac{n+1}{2}}}\subset\widetilde{S}_{V_{i}}=\mathcal{B}^{X_{1\frac{n+1}{2}}}$ for any $i\in\mathcal{Z}_{d_{1}d_{\frac{n+1}{2}}}$.

(ii) The PI sets of $H_{p1,1}^{O}$, $H_{11,\frac{n-1}{2}}^{O}$, $H_{p1,\frac{n+1}{2}}^{O}$, $H_{11,\frac{n+3}{2}}^{O}$, and $H_{p2,1}^{O}$ are $H_{11,\frac{n+3}{2}}^{O}\bigcup \widehat{H}_{2,1}^{O}$, $\widehat{H}_{1,\frac{n+1}{2}}^{O}\bigcup H_{11,\frac{n+3}{2}}^{O}$, $H_{11,\frac{n-1}{2}}^{O}$, $\widehat{H}_{1,1}^{O}\bigcup H_{11,\frac{n-1}{2}}^{O}$, and $\widehat{H}_{1,1}^{O}$, respectively.

(iii) The set sequence is
\begin{equation}
\begin{aligned}
&G_{1}=\widehat{H}_{1,1}^{O}\bigcup H_{11,\frac{n-1}{2}}^{O}\bigcup \widehat{H}_{1,\frac{n+1}{2}}^{O}\bigcup H_{11,\frac{n+3}{2}}^{O},\\
&G_{2}=\widehat{H}_{2,1}^{O}.
\end{aligned}
\end{equation}
For every $H_{p2,1}^{O}\subset G_{2}$, there is $G_{1}\bigcap R_{p2,1}^{O}=\widehat{H}_{1,1}^{O}$. The third condition is true.

(iv) There is a sequence of projection sets $\big\{H_{11,\frac{n-1}{2}}^{O}\rightarrow H_{11,\frac{n+3}{2}}^{O}\rightarrow H_{p1,1}^{O} \big\}^{(X_{1\frac{n+1}{2}})}$. It satisfies that the intersection of two adjacent sets is not empty and their union is the computation basis $\mathcal{B}^{X_{1\frac{n+1}{2}}}$. This means that the family of projection sets $\big\{H_{p1,1}^{O}, H_{11,\frac{n-1}{2}}^{O}, H_{p1,\frac{n+1}{2}}^{O}, H_{11,\frac{n+3}{2}}^{O}, H_{p2,1}^{O}\big\}^{(X_{1\frac{n+1}{2}})}$ is connected.

\section{The proof of theorem 7}\label{G}
According to the Lemma 1 and the symmetry, we only need to consider the POVM on party $X_{1l}$ with $l=2,\ldots,\frac{n}{2}+1$. Let $\widehat{H}_{i,j}^{E}$ represent the union $\bigcup_{p}H_{pi,j}^{E}$.

(I) When $n\geq 10$,
on party $X_{12}$, we choose the subsets $H_{11,1}^{E}$, $H_{p1,\frac{n}{2}}^{E}$, $H_{p1,\frac{n}{2}+1}^{E}$, $H_{21,\frac{n}{2}+2}^{E}$, and $H_{p2,1}^{E}$.

(i) It is not difficult to find that $\mathcal{B}_{i}^{X_{12}}\subset\widetilde{S}_{V_{i}}=\mathcal{B}^{X_{12}}$ for any $i\in\mathcal{Z}_{d_{1}d_{2}}$.

(ii) $H_{11,1}^{E}$ is the PI set of the other subsets $H_{p1,\frac{n}{2}}^{E}$, $H_{p1,\frac{n}{2}+1}^{E}$, $H_{21,\frac{n}{2}+2}^{E}$, and $H_{p2,1}^{E}$. The union of these subsets $\widehat{H}_{1,\frac{n}{2}}^{E}\bigcup \widehat{H}_{1,\frac{n}{2}+1}^{E}\bigcup H_{21,\frac{n}{2}+2}^{E}\bigcup \widehat{H}_{2,1}^{E}$ is the PI set of $H_{11,1}^{E}$.

(iii) For the set sequence
\begin{equation}
\begin{aligned}
&G_{1}=H_{11,1}^{E}\bigcup \widehat{H}_{1,\frac{n}{2}+1}^{E}\bigcup H_{21,\frac{n}{2}+2}^{E}\bigcup \widehat{H}_{2,1}^{E},\\
&G_{2}=\widehat{H}_{1,\frac{n}{2}}^{E},
\end{aligned}
\end{equation}
there is $G_{1}\bigcap R_{p1,\frac{n}{2}}^{E}=H_{11,1}^{E}$. The third condition holds.

(iv) Because the projection set of $H_{11,1}^{E}$ on party $X_{12}$ is the computation basis $\mathcal{B}^{X_{12}}$, the family of projection sets $\big\{H_{11,1}^{E}, H_{p1,\frac{n}{2}}^{E}, H_{p1,\frac{n}{2}+1}^{E}, H_{21,\frac{n}{2}+2}^{E}, H_{p2,1}^{E}\big\}^{(X_{12})}$ is connected.

On party $X_{13}$, we choose the subsets $H_{11,1}^{E}$, $H_{p1,\frac{n}{2}-1}^{E}$, $H_{11,\frac{n}{2}}^{E}$, and $H_{12,1}^{E}$.

(i) Obviously, $\mathcal{B}_{i}^{X_{13}}\subset\widetilde{S}_{V_{i}}=\mathcal{B}^{X_{13}}$ for any $i\in\mathcal{Z}_{d_{1}d_{3}}$.

(ii) $H_{11,1}^{E}$ is the PI set of the other subsets $H_{p1,\frac{n}{2}-1}^{E}$, $H_{11,\frac{n}{2}}^{E}$, and $H_{12,1}^{E}$. The union of these subsets $\widehat{H}_{1,\frac{n}{2}-1}^{E}\bigcup H_{11,\frac{n}{2}}^{E}\bigcup H_{12,1}^{E}$ is the PI set of $H_{11,1}^{E}$.

(iii) For the set sequence
\begin{equation}
\begin{aligned}
&G_{1}=H_{11,1}^{E}\bigcup H_{12,1}^{E},\\
&G_{2}=\widehat{H}_{1,\frac{n}{2}-1}^{E}\bigcup H_{11,\frac{n}{2}}^{E}.
\end{aligned}
\end{equation}
Here $G_{1}\bigcap R_{p1,\frac{n}{2}-1}=G_{1}\bigcap R_{11,\frac{n}{2}}=H_{11,1}^{E}$. The third condition holds.

(iv) Because the projection set of $H_{11,1}^{E}$ on party $X_{13}$ is the computation basis $\mathcal{B}^{X_{13}}$, the family of projection sets $\big\{H_{11,1}^{E}, H_{p1,\frac{n}{2}-1}^{E}, H_{11,\frac{n}{2}}^{E}, H_{12,1}^{E}\big\}^{(X_{13})}$ is connected.

On party $X_{1l}$ ($l=4,\ldots,\frac{n}{2}-1$), we choose the subsets $H_{11,1}^{E}$, $H_{p1,\frac{n}{2}-l+2}^{E}$, $H_{11,\frac{n}{2}-l+3}^{E}$, $H_{p1,\frac{n}{2}+1}^{E}$, and $H_{11,\frac{n}{2}+2}^{E}$.

(i) Obviously, $\mathcal{B}_{i}^{X_{1l}}\subset\widetilde{S}_{V_{i}}=\mathcal{B}^{X_{1l}}$ for any $i\in\mathcal{Z}_{d_{1}d_{l}}$.

(ii) $H_{11,1}^{E}$ is the PI set of the subsets $H_{p1,\frac{n}{2}-l+2}^{E}$, $H_{p1,\frac{n}{2}+1}^{E}$, and $H_{11,\frac{n}{2}+2}^{E}$. The PI sets of $H_{11,1}^{E}$ and $H_{11,\frac{n}{2}-l+3}^{E}$ are the union $\widehat{H}_{1,\frac{n}{2}-l+2}^{E}\bigcup H_{11,\frac{n}{2}-l+3}^{E}$ and $\widehat{H}_{1,\frac{n}{2}+1}^{E}\bigcup H_{11,\frac{n}{2}+2}^{E}$, respectively.

(iii) For the set sequence
\begin{equation}
\begin{aligned}
&G_{1}=H_{11,\frac{n}{2}-l+3}^{E}\bigcup \widehat{H}_{1,\frac{n}{2}+1}^{E}\bigcup H_{11,\frac{n}{2}+2}^{E},\\
&G_{2}=H_{11,1}^{E},\\
&G_{3}=\widehat{H}_{1,\frac{n}{2}-l+2}^{E}.
\end{aligned}
\end{equation}
Here $G_{1}\bigcap R_{11,1}^{E}=H_{11,\frac{n}{2}-l+3}^{E}$ and $R_{p1,\frac{n}{2}-l+2}^{E}=H_{11,1}^{E}=G_{2}$. The third condition is true.

(iv) Because the projection set of $H_{11,1}^{E}$ on party $X_{1l}$ is the computation basis $\mathcal{B}^{X_{1l}}$, the family of projection sets $\big\{H_{11,1}^{E}, H_{p1,\frac{n}{2}-l+2}^{E}, H_{11,\frac{n}{2}-l+3}^{E}, H_{p1,\frac{n}{2}+1}^{E}, H_{11,\frac{n}{2}+2}^{E}\big\}^{(X_{1l})}$ is connected.

On party $X_{1l}$ ($l=\frac{n}{2}$), we choose the subsets $H_{p1,2}^{E}$, $H_{11,3}^{E}$, $H_{p1,\frac{n}{2}+1}^{E}$, $H_{11,\frac{n}{2}+2}^{E}$, and $H_{p3,1}^{E}$.

(i) $\mathcal{B}_{i}^{X_{1l}}\subset\widetilde{S}_{V_{i}}=\mathcal{B}^{X_{1l}}$ for any $i\in\mathcal{Z}_{d_{1}d_{l}}$ because $\big\{H_{p1,2}^{E}\big\}^{(\overline{X_{1l}})}\bigcap \big\{H_{11,3}^{E}\big\}^{(\overline{X_{1l}})}\neq \emptyset$ and $\big\{\widehat{H}_{1,2}^{E}\big\}^{(X_{1l})}\bigcup \big\{H_{11,3}^{E}\big\}^{(X_{1l})}=\mathcal{B}^{X_{1l}}$.

(ii) The PI sets of $H_{p1,2}^{E}$, $H_{11,3}^{E}$, $H_{p1,\frac{n}{2}+1}^{E}$, $H_{11,\frac{n}{2}+2}^{E}$, and $H_{p3,1}^{E}$ are $H_{11,\frac{n}{2}+2}^{E}\bigcup \widehat{H}_{3,1}^{E}$, $\widehat{H}_{1,\frac{n}{2}+1}^{E}\bigcup H_{11,\frac{n}{2}+2}^{E}$, $H_{11,3}^{E}$, $\widehat{H}_{1,2}^{E}\bigcup H_{11,3}^{E}$, and $\widehat{H}_{1,2}^{E}$, respectively.

(iii) For the set sequence
\begin{equation}
\begin{aligned}
&G_{1}=\widehat{H}_{1,2}^{E}\bigcup H_{11,3}^{E}\bigcup \widehat{H}_{1,\frac{n}{2}+1}^{E}\bigcup H_{11,\frac{n}{2}+2}^{E},\\
&G_{2}=\widehat{H}_{3,1}^{E}.
\end{aligned}
\end{equation}
Here $R_{p3,1}^{E}=\widehat{H}_{1,2}^{E}\subset G_{1}$. The third condition is true.

(iv) There is a sequence of projection sets $\big\{H_{11,3}^{E}\rightarrow H_{11,\frac{n}{2}+2}^{E}\rightarrow H_{p1,2}^{E}\big\}^{(X_{1l})}$. It satisfies that the intersection of two adjacent sets is not empty and their union is the computation basis $\mathcal{B}^{X_{1l}}$. This means that the family of projection sets $\big\{H_{p1,2}^{E}, H_{11,3}^{E}, H_{p1,\frac{n}{2}+1}^{E}, H_{11,\frac{n}{2}+2}^{E}, H_{p3,1}^{E}\big\}^{(X_{1l})}$ is connected.

On party $X_{1l}$ ($l=\frac{n}{2}+1$), we choose the subsets $H_{p1,1}^{E}$, $H_{11,2}^{E}$, $H_{p1,\frac{n}{2}+1}^{E}$, and $H_{11,\frac{n}{2}+2}^{E}$.

(i) $\mathcal{B}_{i}^{X_{1l}}\subset\widetilde{S}_{V_{i}}=\mathcal{B}^{X_{1l}}$ for any $i\in\mathcal{Z}_{d_{1}d_{l}}$ because $\big\{H_{p1,1}^{E}\big\}^{(\overline{X_{1l}})}\bigcap \big\{H_{11,2}^{E}\big\}^{(\overline{X_{1l}})}\neq \emptyset$ and $\big\{\widehat{H}_{1,1}^{E}\big\}^{(X_{1l})}\bigcup \big\{H_{11,2}^{E}\big\}^{(X_{1l})}=\mathcal{B}^{X_{1l}}$.

(ii) $\widehat{H}_{1,1}^{E}\bigcup H_{11,2}^{E}$ is the PI set of the subsets $H_{p1,\frac{n}{2}+1}^{E}$ and $H_{11,\frac{n}{2}+2}^{E}$. $\widehat{H}_{1,\frac{n}{2}+1}^{E}\bigcup H_{11,\frac{n}{2}+2}^{E}$ is the PI set of the subsets $H_{p1,1}^{E}$ and $H_{11,2}^{E}$.

(iii) The set sequence is $G_{1}=\widehat{H}_{1,1}^{E}\bigcup H_{11,2}^{E}\bigcup \widehat{H}_{1,\frac{n}{2}+1}^{E}\bigcup H_{11,\frac{n}{2}+2}^{E}$.

(iv) There is a sequence of projection sets $\big\{H_{11,2}^{E}\rightarrow H_{11,\frac{n}{2}+2}^{E}\rightarrow H_{p1,1}^{E}\big\}^{(X_{1l})}$. It satisfies that the intersection of two adjacent sets is not empty and their union is the computation basis $\mathcal{B}^{X_{1l}}$. This means that the family of projection sets $\big\{H_{p1,1}^{E},H_{11,2}^{E}, H_{p1,\frac{n}{2}+1}^{E}, H_{11,\frac{n}{2}+2}^{E}\big\}^{(X_{1l})}$ is connected.

(II) When $n=8$, the proofs on parties $X_{12}$, $X_{13}$, $X_{14}$ and $X_{15}$ are same as the proofs on parties $X_{12}$, $X_{13}$, $X_{1\frac{n}{2}}$, and $X_{1\frac{n}{2}+1}$ in (I), respectively.

(III) When $n=6$, the proofs on parties $X_{12}$ and $X_{14}$ are same as the proofs on parties $X_{12}$ and $X_{1\frac{n}{2}+1}$ in (I).

On party $X_{13}$, we choose the subsets $H_{11,1}^{E}$, $H_{p1,2}^{E}$, $H_{11,3}^{E}$, $H_{p1,4}^{E}$, $H_{11,5}^{E}$, $H_{p3,1}^{E}$ and $H_{4,1}^{E}$.

(i) There is $\mathcal{B}_{i}^{X_{13}}\subset\widetilde{S}_{V_{i}}=\mathcal{B}^{X_{13}}$ for any $i\in\mathcal{Z}_{d_{1}d_{3}}$ because $\big\{H_{p1,4}^{E}\big\}^{(\overline{X_{13}})}\bigcap \big\{H_{11,5}^{E}\big\}^{(\overline{X_{13}})}\bigcap \big\{H_{p3,1}^{E}\big\}^{(\overline{X_{13}})}\neq \emptyset$ and $\big\{\widehat{H}_{1,4}^{E}\big\}^{(X_{13})}\bigcup \big\{H_{11,5}^{E}\big\}^{(X_{13})}\bigcup \big\{\widehat{H}_{3,1}^{E}\big\}^{(X_{13})}=\mathcal{B}^{X_{13}}$.

(ii) For every $H_{pr,t}^{E}$, the corresponding PI set $R_{pr,t}^{E}$ is shown in Table \ref{Gt4}.
\begin{table}[tbp]
\centering
\caption{Corresponding PI set $R_{pr,t}^{E}$ for each subset $H_{pr,t}^{E}$ on the $X_{13}$ party.}\label{Gt4}
\begin{tabular}{cl|cl}
\hline
\hline
Subset~~ & ~~~~~PI set~~~~~ & ~~Subset~~ & ~~~~~PI set~~~~~ \\ \hline
 $H_{11,1}^{E}$ & $R_{11,1}^{E}=H_{11,5}^{E}\bigcup \widehat{H}_{3,1}^{E}$ & $H_{11,5}^{E}$ & $R_{11,5}^{E}=H_{11,1}^{E}\bigcup H_{4,1}^{E}$ \\
 $H_{p1,2}^{E}$ & $R_{p1,2}^{E}=H_{11,5}^{E}\bigcup \widehat{H}_{3,1}^{E}$ & $H_{p3,1}^{E}$ & $R_{p3,1}^{E}=\widehat{H}_{1,2}^{E}$ \\
 $H_{11,3}^{E}$ & $R_{11,3}^{E}=\widehat{H}_{1,4}^{E}$          & $H_{4,1}^{E}$ & $R_{4,1}^{E}=\widehat{H}_{1,4}^{E}\bigcup H_{11,5}^{E}$ \\
 $H_{p1,4}^{E}$ & $R_{p1,4}^{E}=H_{11,3}^{E}$                   & \\
\hline
\end{tabular}
\end{table}

(iii) For the set sequence
\begin{equation}
\begin{aligned}
&G_{1}=\widehat{H}_{1,2}^{E}\bigcup H_{11,3}^{E}\bigcup \widehat{H}_{1,4}^{E}\bigcup H_{11,5}^{E}\bigcup H_{4,1}^{E},\\
&G_{2}=H_{11,1}^{E}\bigcup \widehat{H}_{3,1}^{E}.
\end{aligned}
\end{equation}
Here $R_{11,1}^{E}\bigcap G_{1}=H_{11,5}^{E}$ and $R_{p3,1}^{E}\bigcap G_{1}=\widehat{H}_{1,2}^{E}$. The third condition is true.

(iv) There is a sequence of projection sets $\big\{H_{11,1}^{E}\rightarrow H_{11,5}^{E}\rightarrow H_{4,1}^{E}\rightarrow H_{11,3}^{E}\big\}^{(X_{13})}$. It satisfies that the intersection of two adjacent sets is not empty and their union is the computation basis $\mathcal{B}^{X_{13}}$. This means that the family of projection sets $\big\{H_{11,1}^{E},H_{p1,2}^{E},H_{11,3}^{E},H_{p1,4}^{E},H_{11,5}^{E},H_{p3,1}^{E},H_{4,1}^{E}\big\}^{(X_{13})}$ is connected.

\end{appendix}

\end{document}